\documentclass[12pt]{article}
\usepackage{amsfonts}
\usepackage{amsmath}
\usepackage{epsfig}
\usepackage[square, comma, sort&compress]{natbib}
\def\beq{\begin{equation}}
\def\eeq{\end{equation}}
\def\bea{\begin{eqnarray}}
\def\eea{\end{eqnarray}}
\def\bq{\begin{quote}}
\def\eq{\end{quote}}

\def\tli{\tilde\Lambda^2}

\def\simlt{\stackrel{<}{{}_\sim}}
\def\simgt{\stackrel{>}{{}_\sim}}
\def\Mc{\mathcal{M}}

\begin{document}

\title{
Gravity in Gauge Mediation} 
\author{Z.~Lalak$^1$, S.~Pokorski$^1$  and  K.~Turzy\'nski$^2$\\{\small {}}\\{\small ${}^1${\it Institute of Theoretical Physics, Warsaw University,}}\\{\small {\it ul.\ Ho\.za 69, 00-681 Warsaw, Poland;  }}\\{\small ${}^2${\it Physics Department, University of Michigan,}}\\{\small {\it 450 Church St., Ann Arbor, MI-48109, USA  }}\\}

\date{}
\maketitle
\begin{abstract}
We investigate O'Raifeartaigh-type models for F-term supersymmetry
breaking in gauge mediation scenarios in the presence of gravity.
It is pointed out that the vacuum structure of those
models is such that in metastable vacua gravity mediation contribution to scalar masses is
always suppressed to the level below 1 percent, almost sufficient for avoiding FCNC problem.
Close to that limit, gravitino mass can be in the range 10-100 GeV, opening several interesting
possibilities for gauge mediation models, including Giudice-Masiero mechanism for $\mu$
and $B\mu$ generation.  Gravity sector can include stabilized moduli.

\end{abstract}

\newpage

\section{Introduction}

Gauge mediation of supersymmetry breaking  to the Standard Model sector
\cite{GMfirst01,GMfirst02,GMfirst03,GMfirst04,GMfirst05,GMfirst06,GMfirst07,GMfirst1,GMfirst2,GMfirst3,GMrev} is an interesting alternative to gravity mediation.
A considerable attention has recently been paid to gauge mediation models
with very heavy
messengers and the gravitino mass in the GeV range 
\cite{sspot,Nomura,Nir}, as a phenomenologically interesting possibility.  
Gravitino  in that mass range is  a viable dark matter candidate, consistently  with a high reheating
temperature needed for leptogenesis. 
\cite{Feng0,Feng00,Feng1,Feng15,Feng2,Feng25,Feng3,Feng35,Feng4,Feng5}.
On the other hand,
there has been a revival of interest 
in O'Raifeartaigh-type mechanism of $F$-term supersymmetry breaking
\cite{OR}  as the source of supersymmetry breaking in gauge mediation scenarios, see, e.g.~\cite{Kitano,rozne0,rozne1,rozne2,rozne3,EOGM,rozne4,rozne45,rozne5,rozne6,Carpenter,rozne7}.
This is partly motivated by the wide recognition of the role  of metastable supersymmetry
breaking vacua in models with $R$-symmetry broken spontaneously and/or
explicitly by a small parameter \cite{ISS}. 
With explicit models of supersymmetry breaking in gauge mediation scenarios at hand and 
with special interest in high scale supersymmetry breaking  vacua, it is important to investigate
in detail the role of gravity in such scenarios.

Any physically acceptable model of 
spontaneous symmetry breaking needs gravity for eliminating massless 
goldstino by  the super-Higgs mechanism. But gravity may also be important
for determination of the supersymmetry breaking vacua. A simple example is  a model 
discussed in \cite{Kitano} with the superpotential containing a linear term of 
a gauge singlet chiral superfield $X$ responsible for supersymmetry  
breakdown.  The messengers $Q$ and $q$, transmitting the supersymmetry 
breakdown to the visible sector,     
transform as $\mathbf{5}$ and $\mathbf{\bar 5}$ of $SU(5)$ 
and are coupled to the field $X$. The superpotential reads   
$W = FX-\tilde\lambda XQq$ and the form of the K\"ahler potential,
\begin{equation}
\label{kitanokaehler}
K = \bar{X}X-\frac{(\bar{X}X)^2}{\tli}+\bar{q}q+\bar{Q}Q\, ,
\end{equation}
takes into account the loop corrections representing the logarithmic 
divergence in the effective potential coming from  the effects of the 
massive fields in the O'Raifeartaigh model which have been integrated out. 
The sign of the second term in the K\"ahler potential is negative here.
For $\tilde\lambda=0$ supersymmetry is broken by $F_X\neq0$ and $X$ is 
stabilized at $0$.  Supersymmetry is however restored by turning on 
the coupling $\lambda$. The supersymmetric global minimum is at 
$X=0$ and $Q=q=\sqrt{F/\tilde\lambda}$. Coupling the model to gravity changes 
the vacuum structure. Supersymmetric vacuum is still present as a global 
minimum (with shifted values of the fields) but in addition a local 
(metastable) minimum with broken supersymmetry   
appears, with  vanishing vevs for the messengers fields.

A constant $c$ is added to the superpotential to cancel the 
cosmological constant at the metastable vacuum. 
Another fact worth mentioning is that, with gravity, after decoupling 
the messengers ($\tilde\lambda=0$) the supersymmetric vacuum disappears 
(as in the case without gravity) but the minimum is for $X$ different 
from zero. 
In the above discussion, the role of gravity is linked to the 
negative sign of the second term in the K\"ahler potential, which 
follows from the O'Raifeartaigh model. However, more general models 
of a similar type can give positive sign for that term and $X$ can be 
stabilized away from the origin, with broken supersymmetry \cite{Shih},  
with or without messengers  even in the limit $M_P\to\infty$.  
One can expect that the role of gravity is then  more subtle.

In both cases, the models without messengers can  be used for supersymmetry 
breaking and $F$-term vacuum uplifting in the gravity mediation scenarios \cite{Scrucca1,Scrucca2,Dudas,Scrucca3,OKKLT}.

The purpose of this note is to investigate a certain broad class of 
O'Raifeartaigh-type models (ORT models) \cite{EOGM} as the mechanism 
for supersymmetry breaking in gauge mediation scenarios, coupled to gravity.  The ORT models for 
supersymmetry breaking may contain only SM singlets and then for gauge 
mediation of supersymmetry breaking  they must be supplemented  
with the messenger sector.  The ORT models may themselves contain fields 
charged under SM gauge symmetries, participating in the dynamics of 
supersymmetry breaking and simultaneously transmitting it to the SM. 
Finally, the supersymmetry breaking sector may obey the $SU(5)$ symmetry
or $SU(5)$ symmetry may be explicitly broken by the masses and couplings.
The gravity sector may or may 
not contain additional fields (moduli). 

We investigate the vacuum structure of the globally supersymmetric models and after
coupling them to gravity. We find that in the latter case, metastable local minima with
broken supersymmetry exist
only for the values of the supersymmetry breaking field X stabilized below certain
upper bound of order $10^{-3}M_P$. In addition, at the tree level,  there always exists a stable 
Polonyi-type minimum with $X\approx M_P$ but we are interested only in the vacua with
$X$ stabilized at values such  that quantum corrections from the messenger and O'Raifeartaigh
sectors are under perturbative control.

This result has interesting implications for the possible importance of gravity mediation  
relative to gauge mediation (see also \cite{Dudas2})
in models where gravity 
is coupled to the ORT models of supersymmetry breaking, in the presence 
of  charged under SM gauge group messengers. 
After coupling to gravity, in metastable vacua gauge mediation remains always the dominant
mechanism.  Another version of the same conclusion is that, in moduli stabilization models with $F$-term uplifting in metastable vacua, adding messengers changes gravity mediation
into dominantly gauge mediation.
At the same time, our results show that gauge mediation with messenger masses 
up to $10^{15}\,\mathrm{GeV}$ and gravitino mass up to $100\,\mathrm{GeV}$ is a viable possibility. We investigate
some phenomenological consequences of such scenarios. One question is whether 
the range of values of the $\mu$ and $B\mu$ parameters consistent with the electroweak
breaking  can be compatible with Giudice-Masiero mechanism \cite{GiuMa}.
The idea of employing gravity to solve the $\mu$ problem of gauge mediation
bears some resemblance to the motivation
of \cite{sspot}, though the tools we use and the models we build are
quite different.

We begin in Section \ref{sec2} by reviewing globally symmetric ORT models and 
classifying them according to the existence or nonexistence
of supersymmetry breaking vacuum with $X\neq0$ .
In Section \ref{sec30} we discuss the role of gravity. 
We show that coupling to gravity has important implications for the 
accessible range of values for stabilized $X$.
We also consider the gravity 
sector with moduli fields and discuss the moduli stabilization 
in this context.  
We also review scenarios that, potentially, could lead to 
the dominance of gravity mediation in our present general set-up.
In Section \ref{sec4} we turn our attention to ORT models in which the messenger sector
does not exhibit the unified $SU(5)$ symmetry. Such models have been
advocated in \cite{EOGM} as a possible remedy to the fine-tuning
problem of the MSSM. We study some simple but viable examples
of these models and determine their properties.
We then proceed to Section \ref{sec5} where we discuss the 
phenomenological implications of the models studied in Sections \ref{sec2} and \ref{sec4} 
in the presence of the upper bound on the value of $X$ found in Section \ref{sec3}.
Finally, Section \ref{sec6} contains our conclusions.

\section{Properties of metastable supersymmetry breaking vacua in
generalized O'Raifeartaigh models}
\label{sec2}

In this section we discuss globally supersymmetric ORT models. 
A simple reference model is the original 
O'Raifeartaigh model itself \cite{OR}, 
with its effective form mentioned in the Introduction. We  consider a 
single field $X$ with $R(X)=2$ and with a linear term in the superpotential, 
and a sector consisting of $N$ pairs of fields $\phi_i$, $\tilde\phi_i$ 
belonging to $n_\phi$-dimensional conjugate representations of a 
gauge group.\footnote{For 
self-conjugate representations, we may identify the two sets 
of fields, simultaneously dividing 
$n_\phi$ by 2 in our subsequent formulae. 
In the case of singlets we may also have to impose a parity symmetry 
under which these fields are odd, in order to distinguish 
them from the field $X$.} 
We assume that at tree level these fields have canonical K\"ahler 
potential and that their interactions are described by the most general 
superpotential consistent with $R$ symmetry:
\begin{equation}
\label{or1}
W = FX + \sum_{i=1}^N\sum_{j=1}^N \tilde\phi_i(m_{ij}+\lambda_{ij}X)\phi_j \, ,
\end{equation}
where the mass parameter $m_{ij}$ can be nonzero if and only if 
$R(\tilde{\phi}_i)+R(\phi_j)=2$, 
and the coupling $\lambda_{ij}$ can be nonzero if and only if 
$R(\tilde{\phi}_i)+R(\phi_j)=0$. 
Without loss of generality, we can order the fields so that the 
$R(\phi_j)$ do not increase and $R(\tilde\phi_j)$ do not decrease with 
$j$. 
Interactions of $X$ with $\tilde\phi_i$ and $\phi_j$ induce an effective
potential for $X$. In the leading order in the parameter $F/\bar{m}^2$,
where $\bar{m}$ is a representative mass scale of $m_i$,
the presence of the effective potential can be accounted for by
introducing a correction to the K\"ahler potential \cite{ISS}:
\begin{equation}\label{eff2}
\delta K = -\frac{1}{16\pi^2} \mathrm{Tr}\left[ \Mc^\dagger \Mc \ln\left(\frac{\Mc^\dagger \Mc}{Q^2}\right)\right] \, ,
\end{equation}
where $Q$ is the cutoff of the theory (we take it equal to $M_P$)
and we denote by 
$ \Mc_{ij}=m_{ij}+\lambda_{ij}X$
the $X$-dependent mass matrix for $\tilde\phi_j$, $\phi_i$.
It is known that gaugino masses generated at one loop are proportional
to $F\partial_X\ln\mathrm{det}\, \Mc$. Thus,  in order to 
generate gaugino masses at one loop,
$\mathrm{det} \Mc$ must depend on $X$.
However, in \cite{Shih}, it is shown that for $\mathrm{det}\,m\neq 0$ 
$R$-symmetry implies that $\mathrm{det}\, \Mc=\mathrm{det}\,m$.  
Therefore, in order to generate gaugino masses at one loop,
$ \Mc$ must be singular in the limit $X\to 0$ 
(i.e.~matrix $m$ must have at least one zero eigenvalue) 
\cite{EOGM}.
One possibility of constructing such a model consists in choosing
$R$-charges so that $ \Mc$ can be split into a singular and 
nonsingular part as:
\begin{equation}
 \Mc(X) = \left( \begin{array}{cc} \tilde\lambda X & 0 \\ 0 &  \Mc^\mathrm{ns}(X)\end{array} \right) \, ,
\label{block}
\end{equation}
where the matrix $ \Mc^\mathrm{ns}(X)$ is the maximal 
submatrix of $ \Mc$ such that $\mathrm{det} \Mc^\mathrm{ns}$
does not depend on $X$. We now denote by $\tilde\lambda$ the 
submatrix of $\lambda$ which has not been included in 
$ \Mc^\mathrm{ns}$.

A few comments are in order here. Firstly, the form (\ref{block}) 
is by no means necessary for constructing a successful model of 
supersymmetry breaking and its mediation to the visible sector,
as shown explicitly in \cite{EOGM} and as we show later in Appendix B 
with our 
examples (v) and (vi). However, the form (\ref{block}) is a perfectly 
admissible possibility, dependent only upon the assignment of the 
$R$-charges and we thereby find it interesting to explore the 
structure of such models. The two sectors of the $\tilde\phi_i$ and 
$\phi_j$ fields distinguished by (\ref{block}) need not belong to 
the same gauge representation. Indeed, for generating soft 
supersymmetry breaking masses of the MSSM fields only the sector 
coupled to $\tilde\lambda X$ has to transform nontrivially under the 
gauge group (we shall call it the messenger sector), while it is 
possible that the sector coupling to $ \Mc^\mathrm{ns}$
consists only of gauge singlets. However, the nonsinglets of the 
latter sector (which we shall call the O'Raifeartaigh sector), 
if present,  can give contributions to the masses of the MSSM scalars. 
These contributions are of the order of 
$\sim F\langle X\rangle/(m^\mathrm{ns})^2$,
where $m^\mathrm{ns}$ is a typical mass scale of $ \Mc^\mathrm{ns}$,
so they can (but do not have to) be much smaller than the contribution 
of the first sector which are of the order of 
$\sim F/\langle X\rangle$.

Finally, up to the presence of the messenger sector  
which becomes massless in the limit $X\to0$, 
the models characterized by (\ref{block})
are the Type I models of \cite{EOGM}, and we shall
henceforth refer to them in this manner. 

We shall now discuss the vacuum structure of that class of globally 
supersymmetric models under the additional assumption that  messengers  becoming massless in the limit $X\rightarrow 0$ give negligible 
contribution to the K\"ahler potential (\ref{eff2}).
Thus, we assume that the block $ \Mc^\mathrm{ns}$ in 
(\ref{block}) gives the dominant contribution to (\ref{eff2}).
A global unitary transformation of the fields of the same 
$R$-charge allows writing 
$m_{ij}=m_i\delta_{ij}$ with real and positive $m_i$. 
The case of degenerate mass parameters can be treated as a simple 
limit of a nondegenerate spectrum.
We can then expand the correction (\ref{eff2}) to the 
K\"ahler potential in powers of $|X|$ as:
\begin{equation}
\delta K = -\frac{n_\phi \bar{m}^2}{16\pi^2} \sum_{\ell=0}^\infty f_{2\ell} \cdot \left(\frac{\bar\lambda |X|}{\bar m}\right)^{2\ell} \, .
\label{eff2a}
\end{equation}
In (\ref{eff2a}) we have extracted the overall dependence on the
representative mass scale $\bar{m}$ and coupling strength 
$\bar\lambda$, so that the dimensionless functions $f_{2\ell}$ 
depend only on ratios $\lambda_{ij}/\bar\lambda$ and 
$\rho_i\equiv m_i/\bar m$. Note that the scale $\Lambda$ appearing in
eq.\ (\ref{kitanokaehler}) is now given (for $f_4>0$) by
$\Lambda = 4\pi \bar m/(\bar\lambda^2n_\phi^{1/2}|f_4|^{1/2})$ 
and it can be significantly larger than the scale $\bar m$.
In terms of the coefficients $f_i$, the scalar potential
(with vanishing vevs of the messenger fields)
\begin{equation}
V(\tilde\phi_I,\phi_J, X) = (K_{\bar{X}X})^{-1} |F+\tilde\lambda_{ij}\tilde\phi_i\phi_j|^2 +|X|^2\left( \tilde\lambda_{ij} \tilde\lambda_{ik}^\ast \phi_j\phi_k + \tilde\lambda_{ij}\tilde\lambda_{kj}^\ast\tilde\phi_i\tilde\phi_k^\ast\right)
\end{equation}
reads for $\tilde\phi_i=\phi_j=0$:
\begin{equation}
\label{vgeff}
V(|X|) = F^2 \left[ 1+\frac{n_\phi\bar\lambda^2}{16\pi^2}\left( 4f_4\left(\frac{\bar\lambda|X|}{\bar{m}}\right)^2 + 9f_6 \left(\frac{\bar\lambda|X|}{\bar{m}}\right)^4 + \ldots \right)\right]\, ..
\end{equation}
In this expansion we denoted by $\ldots$ terms of higher order in $|X|$
as well as contributions from the messenger sector. We also 
suppressed the effects of the $|X|^0$ and $|X|^2$ in (\ref{eff2a}).
The former amounts to overall rescaling of the superpotential and the 
latter restores canonical normalization of $X$ by rescaling
$X\to ( 1+(n_\phi\bar\lambda^2f_2)(16\pi^2))X$; 
both are corrections of higher order in powers of 
$\bar\lambda^2/(16\pi^2)$.

Obviously, the vacuum structure in the limit $\tilde\lambda\to0$ 
depends on the signs of the coefficients $f_4$ and $f_6$.  
For $f_4>0$ there is a minimum  for $X=0$ with broken supersymmetry.  
We see the correspondence to the discussion in the Introduction.
Turning on the couplings $\tilde\lambda$  we recover supersymmetric 
minimum at $X=0$. For $f_4<0$  the extremum is for $X\neq0$
and its character depends on the sign of  $f_6$.  Also the value of $X$ in 
the extremum depends of $f_6$.
Turning on the couplings $\tilde\lambda$   
we  get a global supersymmetric minimum at $X=0$, too,   
but for $f_6>0$  the metastable minimum at $X\neq 0$ 
and with  broken supersymmetry is still present.
The presence of a supersymmetry breaking minimum with $X\neq0$
may be endangered by turning on couplings $\tilde\lambda$
in three ways: a large contribution of $\tilde\lambda$ to the scalar 
potential or a development of tachyonic masses in the $X$ 
and/or messenger sector.
None of these situation occurs if, parametrically,
$F/|X|^2<|\tilde\lambda_{ij}|\ll \bar{\lambda}^2|X|/\bar{m}$.
Since the sign of $f_4$ distinguishes between the two significantly 
different possibilities, it is interesting to determine $f_4$ for a 
generic O'Raifeartaigh sector.
Generally, the functions $f_{2\ell}$
depend in a rather complicated way on the original model parameters.
A general method for finding them and explicit results for a somewhat
simplified class of models,
in which one can write 
$\lambda_{ij}=\bar\lambda q_i e^{\imath\varphi_i}\delta_{i+1,j}$,
are presented in Appendix A. 
Using these results, we can calculate $f_4$ and, 
if necessary, $f_6$ in general Type I models, as shown with a few
simple examples in Appendix B.

We have so far assumed that the model is described by the reducible
matrix (\ref{block}). We have  briefly analyzed the necessary 
conditions for the  messenger sector not to affect the position of a 
metastable supersymmetry breaking minimum (if it exists).
Here we would like to  extend the class of models under consideration 
by still keeping the reducible form of the matrix $\Mc$ 
but replacing the submatrix $\tilde\lambda X$ in (\ref{block})  
by
\begin{equation}
 \Mc^\mathrm{s} = \left(\begin{array}{ccccc} \tilde\lambda_1 X & \tilde m_1 & & & \\ & \tilde\lambda_2 X & \tilde m_2 & & \\ & & \ddots & \ddots & \\& & & \tilde\lambda_{\tilde N-1}X & \tilde m_{\tilde N-1} \\ & & & & \tilde\lambda_{\tilde N}X \end{array}\right)
\end{equation}
Writing this expression we assumed that in the messenger sector
all the fields $\phi_i$ have different $R$-charges
and that all the fields $\tilde\phi_j$ have different $R$-charges.
The generalization consists in adding explicit mass terms in the messenger 
sector, but in such a way that in the limit $X\to 0$ the matrix 
$ \Mc^\mathrm{s}$ has at least one zero eigenvalue and the 
gaugino masses can be generated.
In the language of \cite{EOGM}, our model is now a hybrid of a 
Type I and Type II model which is classified there as a Type III model.
From now on we shall assume that all the mass terms $\tilde m_i$ are nonzero, 
otherwise $ \Mc^s$ can be split in two independent blocks 
whose contributions to the effective potential add up. 
We would like to check if the present form of  the messenger sector 
can result in developing a supersymmetry breaking minimum with 
$X\neq0$ despite $f_4>0$.
As before we can extract a representative coupling strength $\bar\lambda'$
and a representative mass scale $\bar m'$ and define coupling and mass
ratios as $q'_i\equiv \tilde\lambda_i/\bar\lambda$ and 
$\rho'_i\equiv \tilde m_i/\bar m$.
Around $X=0$ the eigenvalues of 
$ \Mc^{\mathrm{s}\dagger} \Mc^\mathrm{s}$ 
can then be expanded as
$\bar{m}^2(\rho'_i)^2+\mathrm{O}(\bar{\lambda}'X/\bar{m}')$
and 
$f_0(\bar\lambda|X|/\bar{m})^{2\tilde N}$,
where 
$f_0=\prod_{i=1}^{\tilde N}(q'_i)^2/\prod_{j=1}^{\tilde N-1}(\rho'_j)^2$.
The effective K\"ahler potential (\ref{eff2}) can then be expanded as:
\begin{equation}
\delta K = -\frac{n_\phi\bar{m}^2}{16\pi^2}\left(\ldots+f_4\cdot\left(\frac{\bar{\lambda}|X|}{\bar{m}}\right)^4 + 2\tilde Nf_0 \left(\frac{\bar{\lambda}|X|}{\bar{m}}\right)^{2N}\ln \left(\frac{\bar{\lambda}|X|}{\bar{m}}\right) + \ldots \right]\, .
\label{eff2d}
\end{equation} 
In (\ref{eff2d}) we included the lowest-order nontrivial monomial
(recall that a $|X|^2$ in $\delta K$ only rescales $X$) and the only 
term proportional to $\ln|X|$. For simplicity we also assumed
that the messenger and the O'Raifeartaigh sector are made of the same 
representations. We also set $\bar\lambda'=\bar\lambda$ and $\bar m'=\bar m$ 
assuming that this choice leads to a reliable expansion of $\delta K$.
Now $f_4$ is a sum of two contributions: one from the O'Raifeartaigh
sector and the second coming from the messenger
sector. The latter can also be calculated from our formulae in
Appendix A with 
$q_j$ replaced by $q'_{j+1}$ for $j=1,\ldots,\tilde N-1$ and with 
additional contributions from $q'_{\tilde N}=q'_1$, $\rho'_{\tilde N}=0$
and $\rho'_{\tilde N+1}=\rho_1$. 
The logarithmic term in (\ref{eff2d}) generates  an additional
contribution 
$8\tilde N^3 f_0 (\bar\lambda|X|/\bar m)^{(2\tilde N-2)}\ln(\bar\lambda|X|/\bar m)$ 
to the expansion of the effective potential.
For small values of $|X|$ the negative logarithmic term may compensate a 
positive value of $f_4$ to generate a metastable minimum with $X\neq 0$. 
If $f_4<0$ then both the quartic and the logarithmic term in (\ref{eff2d}) 
pull $X$ away from zero and one should consider higher-order terms to 
find the position of the minimum.
A particular case of this scenario is a possibility that the O'Raifeartaigh 
sector is absent and a metastable supersymmetry breaking minimum arises 
only through interactions of the messenger sector 
(Type II models of \cite{EOGM}).
Some examples of models in which the messenger sector
can affect the position of the supersymmetry breaking minimum
are discussed in Appendix B.

We have so far assumed that the interactions stabilizing
$X$ at the supersymmetry breaking minimum obey a unified $SU(5)$
symmetry (which includes the case of the O'Raifeartaigh sector
consisting of gauge singlets). It has been argued that going
beyond this assumption may lead to interesting phenomenological consequences
(e.g.~higgsino (N)LSP, light gluinos and/or stops, small fine-tuning) 
in both gravity
mediated \cite{martin:comp} and gauge mediated \cite{EOGM,martin:comp2} 
scenarios
of supersymmetry breaking. 
Such models (with gauge mediation) will be discussed in Section \ref{sec4}.

In conclusion, we have discussed in this section globally 
supersymmetric ORT models. 
It is clear  that the vacuum structure of such models depends on their 
details. Both options are open: they may or may  not have metastable 
supersymmetry breaking vacua, with
$X\neq 0$. 

\section{Supergravity vacua of generalized O'Raifeartaigh models}
\label{sec30}

\subsection{Constraints on the scale of supersymmetry breaking}
\label{sec3}
In this section we couple an ORT model to gravity, as an illustration 
of gravity effects on such models. We have checked that those effects 
are generic and our results apply to all models considered here.
They are of two kinds: 
one can be anticipated on the basis of the results of \cite{Kitano}
-- for  ORT models with $f_4>0$, i.e. with no minimum  with broken 
supersymmetry and $X\neq 0$, 
gravity effects lead nevertheless to a metastable minimum of that kind.   
The other gravity effect is universal -- for models with and without 
metastable minimum with $X\neq 0$ in the global limit, 
inclusion of gravity leads to the vacuum structure such that there is an upper bound on the values of $X$ 
in the metastable vacua, $X\ll M_P$.  In addition, 
for the same values of parameters, there always exist a tree-level 
Polonyi-like minimum with
broken supersymmetry and
with $X\approx M_P$. Clearly, those effects have interesting  physical 
implications for the gauge mediation models, to be discussed in the 
next section.

In supergravity, the scalar potential for $X$ is given by:
\begin{equation}
V(X) = e^{K/M_P^2} \left[\left(\partial_XW+(\partial_XK/M_P^2)W\right)\left(\partial_{\bar X}\bar W+(\partial_{\bar X}K/M_P^2)\bar W\right)-3W\bar W/M_P^4\right]\, ,
\end{equation}
where we take $K$ to be the sum of the canonical K\"ahler potential
and the one-loop correction (\ref{eff2}), and we assume that
the superpotential $W$ is (\ref{or1}) plus a constant term $c$
necessary to cancel approximately the cosmological constant.

For $X>0$, the scalar potential can be written in a form 
similar to (\ref{vgeff}):
\begin{eqnarray}
V(X) &=& F^2 \left[ 1-\frac{3c^2}{F^2M_P^2}-\frac{4c}{FM_P^2}X +\right. \nonumber\\
&&+\left.\frac{n_\phi\bar\lambda^2}{16\pi^2}\left(1+\frac{2c}{FM_P^2}X\right)\left(4f_4\left(\frac{\bar\lambda X}{\bar{m}}\right)^2 + 9f_6 \left(\frac{\bar\lambda X}{\bar{m}}\right)^4 + \ldots \right)+\ldots \right]\, ,
\label{vgeff2}
\end{eqnarray}
where the first and second $\ldots$ denote term of higher orders in 
$X$ and $M_P^{-1}$, respectively. The basic difference between the 
scalar potential in supergravity and in the globally supersymmetric 
case (\ref{vgeff}) is the presence of odd powers of $X$ \cite{Kitano}. 
In particular, if the minimum at $X=0$ exist in the global limit, 
i.e.~$f_4>0$, then the linear term always destabilizes it, 
shifting it to a nonzero value.
As an illustrative example  we consider a model admitting both types of 
vacua, depending on the values if its parameters, 
namely, that described in Appendix B, in Example (iii) with $r=1$. 
Upon identification $\bar{m}=m/2$, $\bar\lambda=\lambda$ and $n_\phi=1/2$, 
this model is equivalent to the model with singlet fields and the 
superpotential:
\begin{equation}
W = m\phi_1\phi_3+\frac{R}{2}m\phi_2^2+\lambda \phi_1\phi_2 X +FX+c\, .
\label{wshih}
\end{equation}
A model described by (\ref{wshih})  has been extensively studied in 
the global limit in \cite{Shih}. Here we couple it to gravity
 and we again assume that the messenger sector does not affect the 
position of the local supersymmetry breaking minimum.
When $c=FM_P/\sqrt{3}$ the effective potential (\ref{vgeff2})
vanishes in the limit $|X|\to 0$.
Only small corrections to this relation are necessary to make the
cosmological constant vanish at the supersymmetry breaking local minimum 
of the potential with $X\neq0$ and we shall use this
approximate relation from now on.

Functions $f_4$ and $f_6$, defined in eq.\ (\ref{eff2a}), 
have the following form:
\begin{eqnarray}
f_4 &=& -\frac{1+2R^2-3R^4+R^2(R^2+3)\ln R^2}{(R^2-1)^3} \\
f_6 &=& \frac{1+27R^2-9R^4-19R^6+6R^2(R^4+5R^2+2)\ln R^2}{3(R^2-1)^5} \, .
\end{eqnarray}
As shown in \cite{Shih}, the function $f_4$ is positive for $R<2.11$ 
and negative otherwise (see also Figure \ref{fex3} in Appendix B). 
The function $f_6$ is positive for $R>1/2$, thereby ensuring
(in the global limit) the existence of a metastable supersymmetry 
breaking minimum whenever $f_4<0$. 

\begin{figure}
\begin{center}
\includegraphics*[height=5cm]{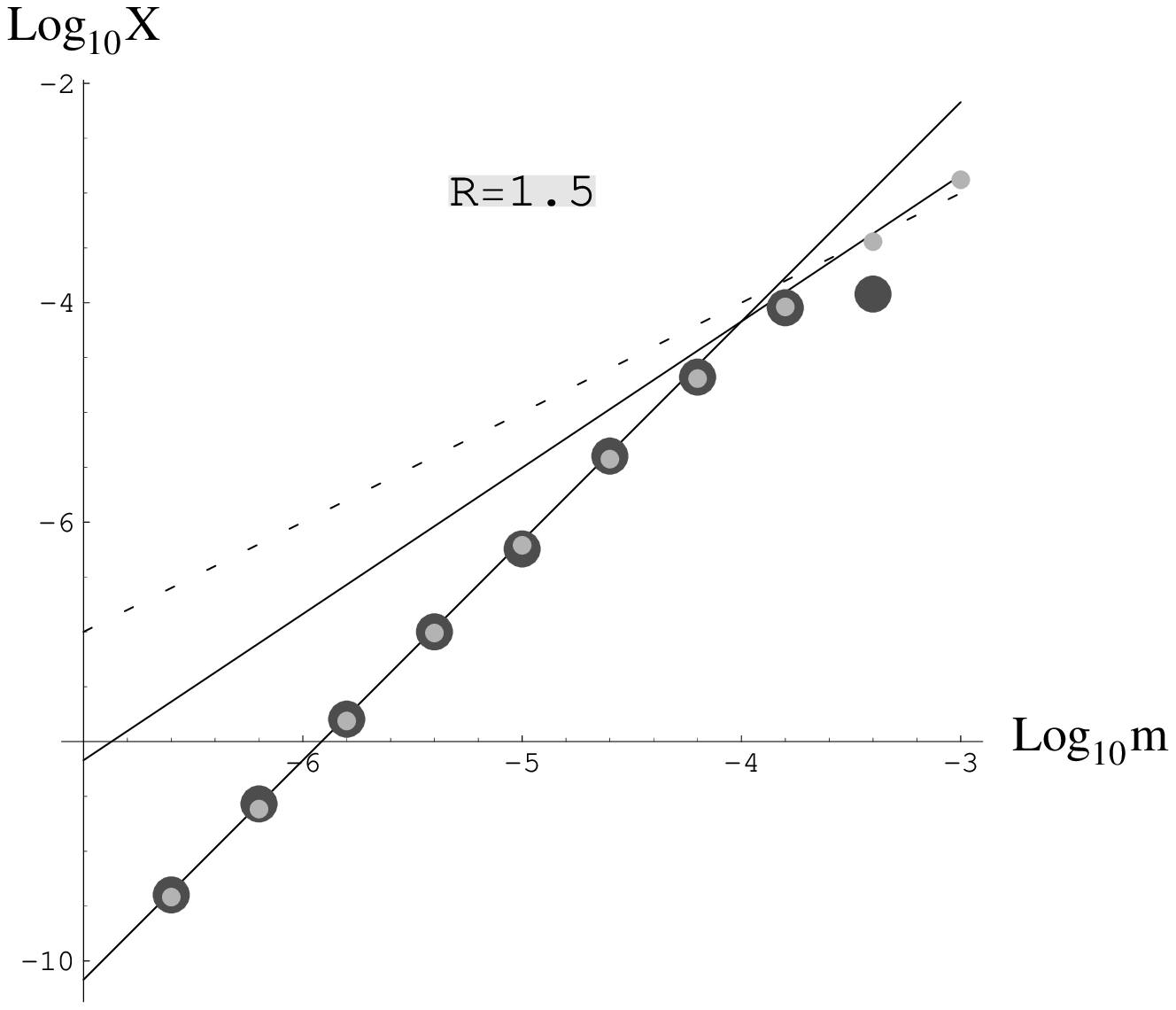}
\hspace{0.3cm}
\includegraphics*[height=5cm]{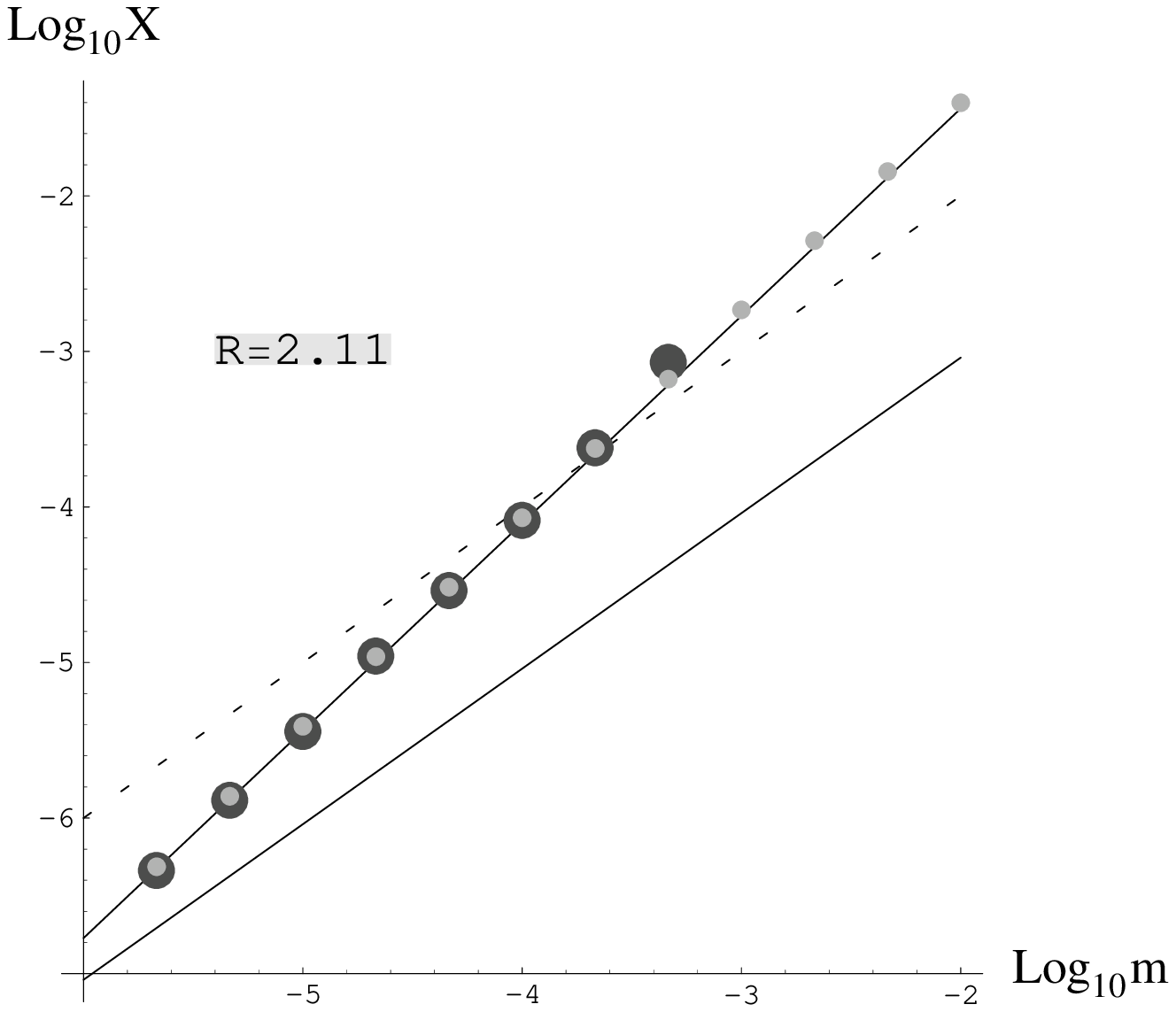}
\hspace{0.3cm}
\includegraphics*[height=5cm]{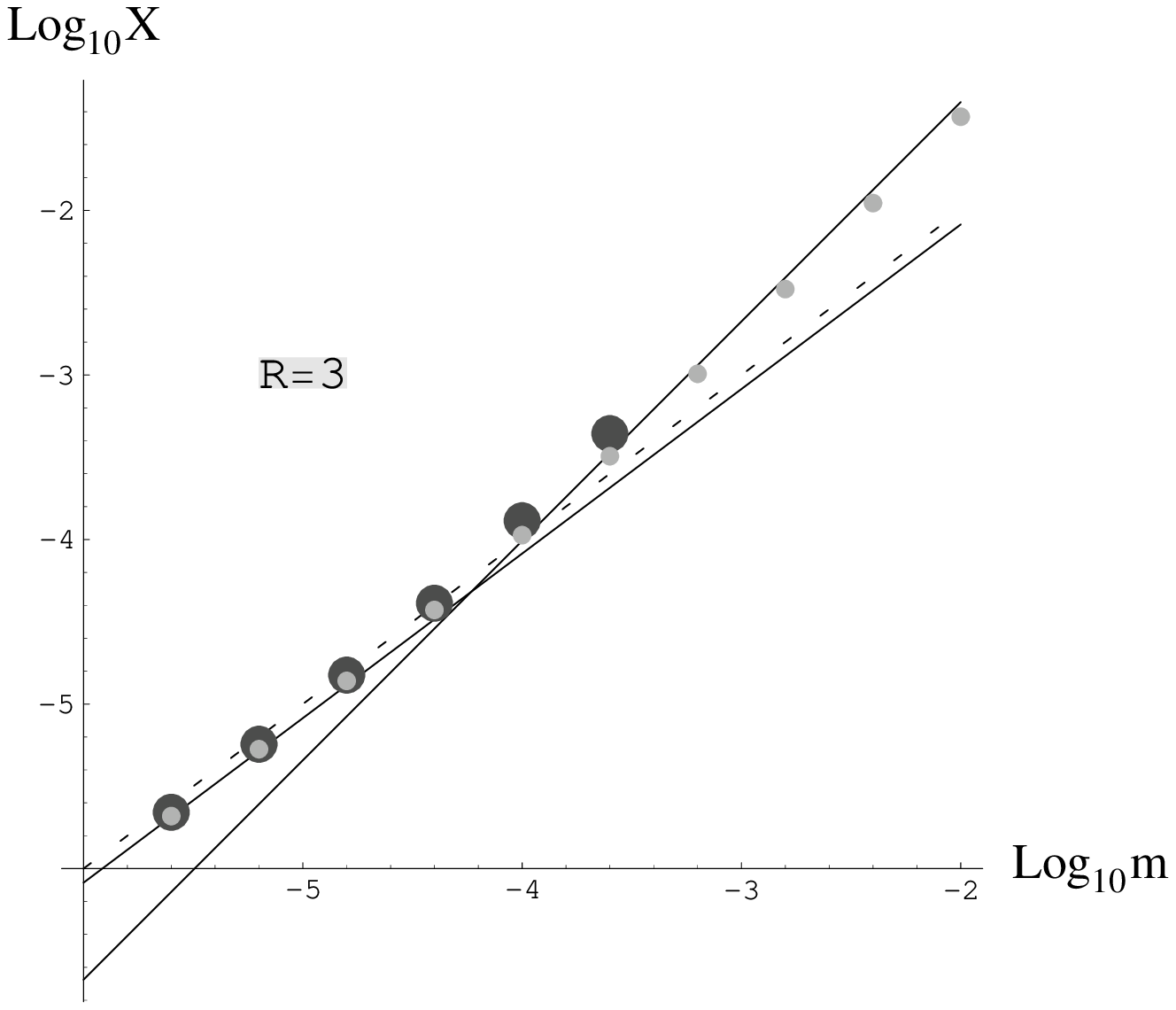}
\end{center}
\caption{\em Positions of the minimum of the full supergravity potential
for $R=1.5$, $R=2.11$ and $R=3$ shown as large dark dots (full calculation),
small light dots ($\delta K$ truncated at $\mathrm{O}(|X|^6)$)
and solid lines (approximate solutions described in the text) 
for $\lambda=1$. The dashed lines show $\lambda X/m=1$. \label{fmphi}}
\end{figure}

If the approximation in the effective potential shown in (\ref{vgeff2})
is sufficiently accurate to determine the local supersymmetry
breaking minimum with $X\neq0$, we can envision three main classes of 
solutions depending on the sign and size of $f_4$.
For $f_4<0$ and $f_6>0$ we recover the minimum previously discussed
in the context of globally supersymmetric model:
\begin{equation} 
X^2 = \frac{8|f_4|}{9f_6}\frac{\bar m^2}{\bar \lambda^2} \, .
\label{ssol1}
\end{equation}
For $f_4>0$ the position of the supersymmetry breaking minimum is 
determined by a balance between terms linear and quadratic in $X$
\cite{Kitano}. The solution reads then:
\begin{equation}
X = \frac{1}{2\sqrt{3}}\frac{\Lambda^2}{M_P} \, ,
\label{ksol}
\end{equation}
where, as before, $\Lambda$ appearing in (\ref{kitanokaehler})
is given by $\Lambda = 4\pi \bar m/(\bar\lambda^2n_\phi^{1/2}|f_4|^{1/2})$.
Finally, we may have $f_4\approx 0$, leading to a dominance of the
quartic term over the quadratic one.
We find then:
\begin{equation}
X^3  =  \frac{16\pi^2}{9\sqrt{3}n_\phi f_6} \frac{\bar m^4}{\bar \lambda^6M_P}\, .
\label{ssol2}
\end{equation}
Note that including the supergravity corrections to the effective
potential is crucial for the existence of
solutions (\ref{ksol}) and (\ref{ssol2}),
for which $\langle X\rangle$ is proportional to negative powers of $M_P$. 
All the three solutions (\ref{ssol1})-(\ref{ssol2})
break $R$-symmetry \cite{Nelson}: 
in (\ref{ssol1}) $R$-symmetry is broken spontaneously, whereas the form of 
(\ref{ksol}) and (\ref{ssol2})
shows that explicit soft $R$-symmetry breaking (the constant term $c$ in 
the superpotential) is transmitted to $\langle X\rangle$ through 
gravitational interactions.
In Figure \ref{fmphi}, we present results of numerical analysis
for $R=3$, $R=2.11$ and $R=1.5$,
basically corresponding to the solutions 
(\ref{ssol1}), (\ref{ssol2}) and (\ref{ksol}), respectively,
in terms of the original parameter $m$ in (\ref{wshih}).
We conservatively assumed a rather large value $\lambda=1$.
Large dark dots represent the position of the minima
calculated in the full supergravity potential with $\delta K$
defined in (\ref{eff2}) and with the cosmological constant
vanishing at the minimum. 
Small light gray dots correspond to minima
calculated in the full supergravity potential, 
but with $\delta K$ truncated at $\mathrm{O}(|X|^6)$ terms. 
Pairs of solid lines show the solutions (\ref{ssol1}) and (\ref{ssol2}) 
for $R=3$ and $R=2.11$, and the solutions (\ref{ksol}) and (\ref{ssol2})
for $R=1.5$. The dashed lines show $\lambda X/m=1$, i.e.\ the value of $X$
for which the expansion in (\ref{eff2a}) is expected to break down.
Our numerical analysis shows that all the three solutions
can be realized in the simple model (\ref{wshih}),
depending on the value of the mass ratio $R$.
We can also see that for increasing values of 
$m$ (and thus $\langle X\rangle$)  
the quartic term in (\ref{vgeff2}) 
becomes more important and for given $R$ solutions 
(\ref{ssol1}) and (\ref{ksol}) can be replaced by (\ref{ssol2}). 
However, unless $|f_4|\ll f_6$, this regime occurs rather close to the 
$\lambda X/m=1$.

We note that the local minimum disappears for mass scales
of the  O'Raifeartaigh sector slightly smaller than $10^{-3}M_P$
This can be understood by noticing that the solution (\ref{ssol2}),
which is a good approximation for sufficiently large $\bar m$,
can be rewritten as:
\begin{equation}
\left(\frac{\bar\lambda X}{\bar m}\right)^3 = \frac{32\pi^2}{9\sqrt{3}f_6}\frac{\bar m}{\bar\lambda^3 M_P^3}
\label{ssol22}
\end{equation}
If the left-hand side of (\ref{ssol2}) exceeds unity, our perturbative 
expansion (\ref{eff2a}) breaks down and one should not expect the minimum
to persist. It also follows from (\ref{ssol22}) that the maximal
scale $\bar{m}$ for which there exists a local minimum with $X\neq0$
scales as $\bar\lambda^3$, which, upon substitution to (\ref{ssol2}) 
shows that the corresponding value of $X$ scales as $\bar\lambda^2$.
These observations are confirmed by our numerical analysis.
Hence our example with $\bar\lambda=1$ is
a conservative choice indeed. 

We repeated the numerical analysis described above for several randomly
generated models with 
the O'Raifeartaigh sector consisting from
$N=4$ to $N=10$ pairs of fields 
(for the details of the randomization procedure,
see Example (iv) in Appendix B)
and checked if
the metastable minimum disappears for large $\bar m$. We found that
this is the case, indeed; in each example we had $X/M_P<O(10^{-3})$,
which shows that the results for the particular example discussed
in this Section can be applied to a broad class of ORT models, including
those with broken $SU(5)$ symmetry, to be discussed in Section \ref{sec4}. 
This bound can, however, be made weaker
by introducing a messenger and/or O'Raifeartaigh sector with a very large
number of fields, so that the one-loop corrections can compete
more efficiently with $M_P$-suppressed corrections.

We also note that although the three possibilities discussed above form an 
exhaustive list for the simple model we have considered, 
in more general situations one can have 
$f_4<0$ and $f_6<0$, and to determine the position
of the local minimum of the potential (\ref{vgeff}) or (\ref{vgeff2})
-- if it exists -- one has to consider terms of higher powers of $X$.
Since in this case one obtains a solution that survives 
in the limit $M_P\to\infty$,
we believe that it has similar properties to (\ref{ssol1}).

The numerical analysis supports the conclusion that  the 
supergravity corrections
provide an upper bound on the values of the
mass scale $\bar m$ of the O'Raifeartaigh sector for which
the metastable supersymmetry breaking minima exist.
In the particular example analyzed here,this bound is 
$\bar m\simlt 10^{-3}M_P$, which corresponds to 
$\Lambda\simlt 10^{-2}M_P$. 

\subsection{Gauge mediation in the presence of moduli}
\label{sec32}

In this subsection we would like to extend our discussion
to gauge mediation in the presence of moduli or, vice versa,
to the impact of messengers on moduli stabilization.
The presence of moduli is a generic future of gravitational 
sectors rooted in extra-dimensional models.  
In fact, ORT models,  with SM singlets only,  have been used as the source of 
supersymmetry breaking and the uplifting potential  
in supergravity models with moduli, stabilized by 
supersymmetric dynamics involving fluxes and nonperturbative 
superpotentials \cite{Scrucca1,Dudas,Scrucca3,OKKLT}.
In simple models discussed in the literature, 
one assumes that all moduli except one
(let us call it $T$) are stabilized by fluxes and 
$T$ is stabilized by a nonperturbative superpotential. 
Its  K\"ahler potential reads $K'=-3\ln(T+\bar T)$ and a 
general superpotential $W^c(T)$ involves a constant term and nonpertubative  
term, such as that of the KKLT model \cite{KKLT},
$W^c(T)=W_0+Ae^{-aT}$, or the KL model $W^c(T)=W_0+Ae^{-aT}-Be^{-bT}$
\cite{KL}. The minimization of the supergravity potential for the modulus 
in the absence of the O'Raifeartaigh sector amounts to solving 
$D_TW=0$, where $D_TW=W_T-(3/(T+\bar T))(W/M_P^2)$,.
i.e.~the minimum with $T$ stabilized at $T_0$ is supersymmetric  and in AdS.

It is rather clear (see e.g.~\cite{Dudas}) that the shift $\delta T$
induced by the O'Raifeartaigh sector is small, provided the 
supersymmetry breaking parameter $F$ in eq.\ (\ref{or1}) is $F\ll M_P^2$ 
(we assume here that the parameters in the superpotential (\ref{or1})
do not depend on the values of the moduli). 
Furthermore, the induced supersymmetry breaking in the modulus sector 
is small. $F_T\ll F_X$.
Hence, in such  simple models the presence of the modulus sector
stabilized at the tree level does not interfere with supersymmetry 
breaking in the ORT models, even if gravitationally induced
like in \cite{Kitano} or in models with $f_4>0$ discussed in Section \ref{sec2}.  
The moduli sector replaces the constant $c$ in the potential (\ref{wshih}), 
to fine-tune the cosmological constant to zero.

It is worth checking the constraints on 
the moduli sector in the presence of messengers in the 
ORT models coupled to gravity with stabilized moduli.  
The scalar potential now reads (for real parameters $A$, $W_0$ and for $X>0$):
\begin{eqnarray}
V(X,T) &=& \frac{F^2M_P^3}{(T+\bar T)^3} \left[ 1-\frac{(T+\bar{T})^2 |W^c_T|^2}{3F^2M_P^2} -\ldots -\frac{4W^c}{FM_P^2}X + \right. \nonumber\\
&& \left.+\frac{n_\phi\bar\lambda^2}{16\pi^2}\left(1+\frac{2W^c}{FM_P^2}X\right)\left(4f_4\left(\frac{\bar\lambda X}{\bar{m}}\right)^2 + 9f_6 \left(\frac{\bar\lambda X}{\bar{m}}\right)^4 + \ldots \right)+\ldots \right]\, ,
\label{vgeff3}
\end{eqnarray}
where we assumed the validity of supergravity approximation $aT\gg 1$
and denoted by $\ldots$ terms subleading in powers of $aT$.
The new element of the present setup is the constraint following from the cancellation of the cosmological constant, correlating the value of  the  $F$ parameter in the superpotential (\ref{or1})
with the value of $T$ in the presence of the dominantly gauge mediated soft terms. 

\begin{figure}
\begin{center}
\includegraphics*[height=7cm]{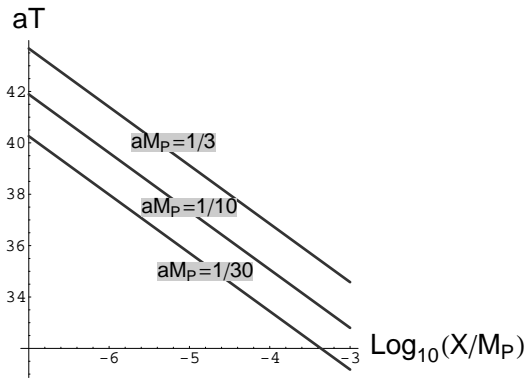}
\end{center}
\caption{\em Values of $aT$ following from relation (\ref{ubt}) 
as a function of $\langle X\rangle/M_P$ for three values of 
$aM_P=1/3,\,1/10,\,1/30$. \label{fmod1}}
\end{figure}

In the KKLT model uplifted by the O'Raifeartaigh sector \cite{OKKLT}, 
the condition of the cosmological constant cancellation 
in the limit $X\to0$, following
from (\ref{vgeff3}), gives
\begin{equation}
\frac{2}{\sqrt{3}M_P} A aT e^{-aT} = F \, ,
\label{tenum}
\end{equation} 
(again corrections from the fact that the actual minimum is located
at $0<\langle X\rangle\ll M_P$ are negligible).
Also, since both the supersymmetric masses squared and supersymmetry breaking
mass squared splitting within multiplets are rescaled by a factor 
$e^{K'}(T+\bar T)^{-3}$ both the gravitino mass $m_{3/2}$
and the soft masses 
have to be multiplied by a factor $e^{K'/2}=(T+\bar{T})^{-3/2}$. 
With this correction, using  equation 
(\ref{tenum}) and relation $m_\mathrm{gaugino}=(\alpha/4\pi)(F/\langle X\rangle)$
(cf.\ eq.\ (\ref{softabove}) in Section \ref{sec5})
and noting that the coupling to the modulus does not alter the 
value of $X$ in the minimum
we obtain the following lower bound on the stabilized modulus $T$:
\begin{equation}
(aM_P)^{3/2} \frac{e^{-aT}}{\sqrt{aT}} = 10^{-14} \left(\frac{m_\mathrm{gaugino}}{100\,\mathrm{GeV}}\right) \left(\frac{0.04}{\alpha}\right) \left(\frac{M_P^3}{A}\right) \frac{\langle X \rangle}{M_P} \, .
\label{ubt}
\end{equation}
For the reference values given above, 
the value of the modulus, $T$, following from this
relation is shown in Figure \ref{fmod1} as a function of 
$aM_P$ for two values of $\langle X\rangle/M_P=10^{-3},\,10^{-6}$. 
Changing
the right-hand side of (\ref{ubt}) by a factor of a few does not result in
any significant changes of the value $T$.
These rather large values of $T_\mathrm{min}$ reflect, in part, 
the smallness of the ratio (\ref{justabove}) in ORT models
requiring the modulus vev larger by 
$\sim\frac{1}{a}\ln(m_\mathrm{gaugino}/m_{3/2})$ 
with respect to usual scenarios with gravity mediation \cite{OKKLT}.
The mass scale of $X$ depends on which supergravity solution
is realized: for (\ref{ssol1}) we can estimate:
$m_X \sim O(1)\,\mathrm{TeV}\, (m_\mathrm{gaugino}/100\,\mathrm{GeV})(0.04/\alpha)(\bar\lambda/1)$, while for (\ref{ksol}) we find
$m_X \sim O(1)\,\mathrm{TeV}\, (m_\mathrm{gaugino}/100\,\mathrm{GeV})(0.04/\alpha)\sqrt{10^3\langle X\rangle/M_P}$.
The mass scale of the modulus is
$m_T \sim (aT)m_{3/2}\sim (2/3)M_P(A/M_P^3)(aM_P)^{3/2}(aT)^{1/2}e^{-aT}$; 
a requirement that so estimated $m_T>100\,\mathrm{GeV}$
can be translated to $aT<34,\,35.5,\,37.5$ for $aM_P=1/30,\,1/10,\,1/3$,
respectively.
Depending on cosmological history
such a light, only gravitationally coupled scalar is prone to
introduce the well-known cosmological moduli problem.

We note that one can attempt to loosen the correspondence
between the parameter $F$ 
and the gravitino mass,
thereby relaxing the bounds discussed above.
An example of such a strategy consists in coupling the
messengers to a fields that plays only a secondary role
in supersymmetry breaking, which suppresses the gauge mediated
contributions to soft masses (see also \cite{oli}). 
The minimal possibility of 
this type utilizes
$F_T\ll F_X$ and employs a superpotential coupling 
$\tilde\lambda e^{-bT}Qq$ instead of $\tilde\lambda XQq$.
This leads to the dominance of gravity mediation, introducing
potentially large FCNC effects, or, if gravity mediation can
be suppressed by means of sequestering, the dominance
of the anomaly mediation giving negative masses squared to
sleptons, though the latter problem may potentially be cured
by $\sim (\alpha/4\pi)(F_X\langle X\rangle/\bar m^2)$
contributions to the scalar soft masses, mediated
by the gauge interactions of nonsinglets in the 
O'Raifeartaigh sector. 
Instead of coupling messengers only to moduli
one can also extend the model by adding another singlet $X'$
with a superpotential $F'X'$ with $F'\gg F$. 
The field $X'$ can be stabilized
at a supersymmetry breaking minimum with an O'Raifeartaigh
sector of its own or by one shared with the field $X$.
Both scenarios go beyond the scope of this work
and we shall leave them for a future study.

Various  mass relations discussed above result from the fact
that with $aT\gg 1$, the modulus sector has practically one
mass scale tightly connected to $F$ through cancellation of the
cosmological constant. More general models, e.g.~the KL model \cite{KL}
obtained by adding another exponential term to the superpotential,
allow for a complete separation of the modulus and messenger sector
and the above conclusions following from (\ref{tenum})
no longer hold.

\section{Models with broken $SU(5)$ symmetry}
\label{sec4}

We have so far discussed models whose messenger sectors obey
the unified $SU(5)$ gauge symmetry. Motivated by \cite{EOGM},
where breaking of $SU(5)$ symmetry allowed constructing models
of gauge mediation
with a low messenger scale and a small $\mu$ parameter,
we would like to extend this discussion to models with an arbitrary messenger
scale. Here we review the properties of such simple models,
concentrating on the predictions which
depend on the structure of these models rather than on the actual model
parameters. This discussion is then continued in Section \ref{sec5}, where we calculate
the low energy spectra of such models and discuss their consequences.

We begin by recalling 
the general formulae for the soft masses in gauge mediated
models. Using the Giudice-Rattazzi method of extracting the supersymmetry
breaking effects from wave function renormalization \cite{GRwave} we find
the one loop gaugino masses \cite{EOGM}:
\beq
\label{mass:gaugino}
M_r = \frac{g_r^2}{16\pi^2} F\,\partial_X\ln\mathrm{det}\,\Mc
\eeq
and the two-loop scalar masses:
\beq
\label{mass:scalar}
m_{\tilde f}^2 = \sum_{r=1}^3 (m_{\tilde f}^{(r)})^2 \, ,\qquad\textrm{with}\qquad
(m_{\tilde f}^{(r)})^2 = \frac{g_r^4}{128\pi^4}C_{\tilde f}^{(r)}F\,\partial_{\bar X}\partial_X \left[\sum_{i=1}^N \ln^2\left(\frac{\Mc_i^2}{\Lambda_\mathrm{UV}^2}\right)\right]\, ,
\eeq
where $r$ runs over the Standard Model gauge groups, $C_{\tilde f}^{(r)}$ are
the appropriate Casimir invariants and $\Mc_i^2$ are the eigenvalues of the
mass matrix $\Mc^\dagger \Mc$. 
Equations (\ref{mass:gaugino}) and 
(\ref{mass:scalar}) are valid at the scale at which we integrate out the
messenger fields\footnote{Of course, the messenger masses can be split; 
corrections to (\ref{mass:scalar}) arising from messenger mass hierarchies
are briefly discussed in Appendix C.}, 
below this scale a usual renormalization group analysis
is required.
Since the Casimir invariants are the same above and below the messenger
threshold, no $A$-terms are generated at the messenger scale.
Equations (\ref{mass:gaugino}) and 
(\ref{mass:scalar}) also assume universal interactions within each messenger
multiplet ($\Mc_i^2$ do not depend on $r$); 
if this assumption is violated, e.g.~when doublets of $SU(2)$ and
triplets of $SU(3)$ forming a $\mathbf{5}$ representation of $SU(5)$
have different mass matrices, we should replace $\Mc_i^2$
by their appropriate components. 

In $SU(5)$-symmetric 
models with $N_\mathrm{mess}$ pairs of messengers in 
$\mathbf{5}+\mathbf{\bar{5}}$
representations and 
with supersymmetric masses proportional to $\langle X\rangle$, 
it follows from (\ref{mass:gaugino}) and (\ref{mass:scalar})
that the soft gaugino and scalar masses obey:
\beq
(m_{\tilde f}^{(r)})^2 = \frac{2C_{\tilde f}^{(r)}}{N_\mathrm{mess}} M_r^2 \, .
\eeq
In models with more complicated messenger interactions, 
in particular in those with broken $SU(5)$ symmetry,
it is convenient to define, by analogy, an {\sl effective messenger number} \cite{EOGM}:
\beq
\label{neffdef}
N_{\mathrm{eff},r} = \frac{2 C_{\tilde f}^{(r)}M_r^2}{(m_{\tilde f}^{(r)})^2} \, ,
\eeq
where $M_r$ and $(m_{\tilde f}^{(r)})^2$ are given by (\ref{mass:gaugino})
and (\ref{mass:scalar}), respectively.
Of course, in this definition we only take these scalars for which 
$(m_{\tilde f}^{(r)})^2\neq0$; the definition is independent of the 
specific choice of $\tilde f$. With messengers residing in 
$\mathbf{5}+\mathbf{\bar{5}}$ pairs 
(with possibly different interactions of the doublet
and triplet components), the effective messenger numbers obey
$5/N_{\mathrm{eff,1}}=3/N_{\mathrm{eff,2}}+2/N_{\mathrm{eff,3}}$.

\begin{figure}
\begin{center}
\includegraphics*[height=7cm]{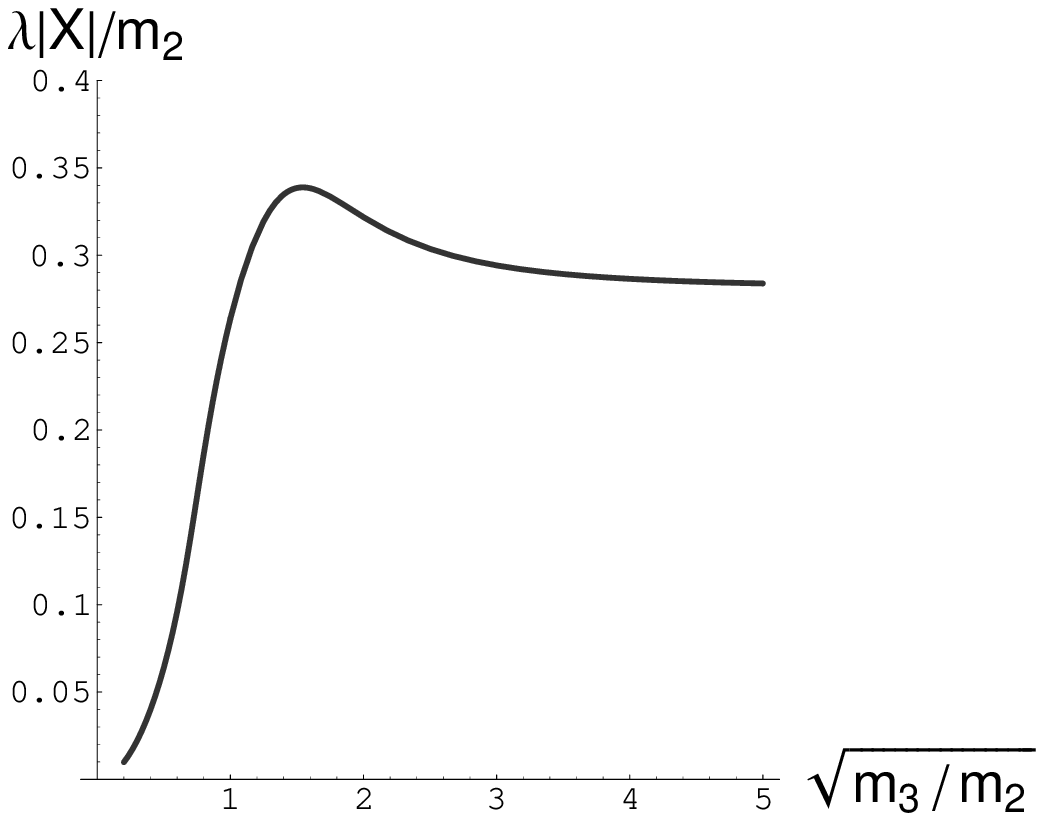}
\hspace{0.5cm}
\includegraphics*[height=7cm]{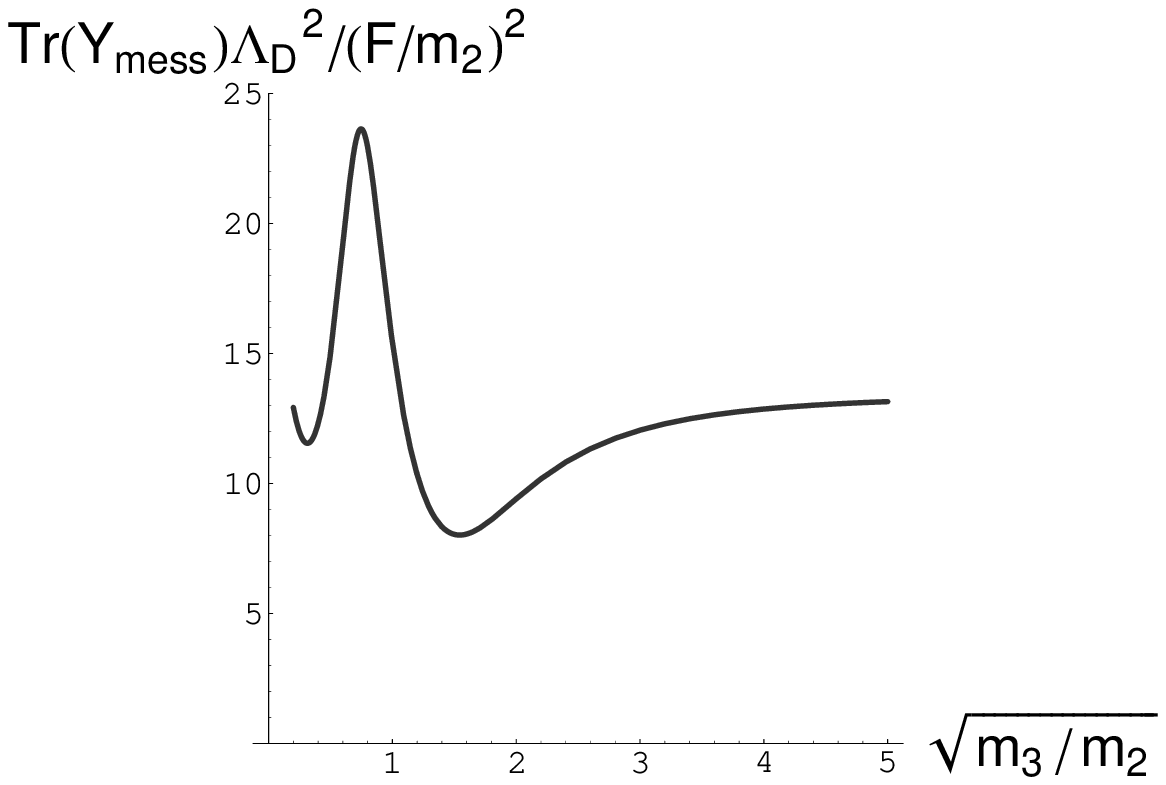}
\end{center}
\caption{\em The location $X$ of the supersymmetry breaking minimum 
and the parameter $\mathrm{Tr}[Y_\mathrm{mess}\Lambda_D]$
describing the one-loop contributions to scalar masses
as functions of the ratio of the mass parameters in the doublet
and triplet sectors, $m_2$ and $m_3$,
for the model defined in (\ref{badmodel}). 
\label{fdt2}}
\end{figure}

It has long been known that
breaking the $SU(5)$ symmetry generically leads to one-loop
$D$-term contributions to the masses of scalars, proportional
to their hypercharge and thus not positive definite \cite{GMfirst2},
unless the messenger sector obeys so-called
{\sl messenger parity} \cite{Dvali}.
The absence of one-loop contributions to scalar masses
is thus a nontrivial constraint on the models. We illustrate this by
studying  this issue in the simplest version of the 
model put forward in an early version of \cite{EOGM}, which
employs three pairs of $\mathbf{5}+\mathbf{\bar 5}$ fields whose
interactions are described by the following superpotential:
\beq
W = FX+ \lambda X(\tilde\phi_1\phi_1+\tilde\phi_2\phi_2) + \lambda' \tilde\phi_3\phi_3 +m_A \tilde\phi_1\phi_2 +\delta m_A \tilde\phi_2\phi_3 \, ,
\label{badmodel}
\eeq 
where $m_2$ and $\delta m_2$ ($m_3$ and $\delta m_3$) are the mass
parameters for the doublets (triplets) residing in 
$\mathbf{5}+\mathbf{\bar 5}$. Following \cite{EOGM}, we take
$\lambda'=\lambda/10$, $\delta m_2=m_2/10$ and $\delta m_3=0$.
We then solve numerically for the minimum of the effective
potential and we calculate the one-loop contributions to the scalar
masses, which in the leading order in $F$ is given by \cite{Dimopoulos}:
\beq
(m_{\tilde f}^2)_\mathrm{1-loop} = \frac{\alpha_1}{4\pi} Y_{\tilde f} \mathrm{Tr}[Y_\mathrm{mess} (\Lambda_D)^2] \, .
\eeq
In this formula, the
trace runs over all the Standard Model states forming
the messenger multiplet
and the parameter $(\Lambda_D^R)^2$
for the states in the SM representation $R$
($SU(2)$ doublets and $SU(3)$ triplets)
is given by:
\beq
(\Lambda_D^R)^2 = \frac{1}{2} \sum_{i,j} \frac{|F^{(R)}_{ij}|^2-|F^{(R)}_{ji}|^2}{(\Mc_i^{(R)})^2} g\left(\frac{(\Mc_j^{(R)})^2}{(\Mc_i^{(R)})^2}\right) \, ,
\label{LambdaD}
\eeq
where $(\Mc^{(R)}_i)^2$ are the eigenvalues of $\Mc^\dagger \Mc$ 
for different representations $R$
and
$F^{(R)}_{ij}$ are the supersymmetry breaking masses of the fields $\tilde\phi_i$
and $\phi_j$ in the scalar potential written in the basis in which
the supersymmetric mass matrix $\Mc$ is diagonal.
The scalar potential for the components of the messenger fields 
in representation $R$ of the SM gauge group then reads
$V=\sum_i (\Mc^{(R)}_i)^2(\phi_i^{(R)\ast}\phi_i^{(R)}+\tilde\phi_i^{(R)\ast}\tilde\phi_i^{(R)})+\sum_{i,j} (F^{(R)}_{ij} \tilde\phi_i^{(R)}\phi_j^{(R)}+\mathrm{h.c.})$.
The function $g$ is given by $g(x)=2/(1-x)+(1+x)/(1-x)^2\ln x$.
In Figure \ref{fdt2}, we show,
as a function of $\sqrt{m_3/m_2}$,  the results for $\lambda |X|/m_2$ at the minimum
and for $\mathrm{Tr}[Y_\mathrm{mess} (\Lambda_D)^2]$ normalized to
$(F/m_2)^2$ which represents a typical mass scale (squared) in usual
expressions for gauge mediation. From this we can estimate that
the ratio of the two-loop contributions to one-loop ones is
$\sim(1/(4\pi))(\alpha_r^2\alpha_1)(1/10)$ which is an unavoidably
small number. Since these one-loop contributions are dominant but 
not positive definite for right-handed sleptons, their presence
is phenomenologically unacceptable.

\subsection{Models with $1\approx N_{\mathrm{eff},2}<N_{\mathrm{eff},3}$}
\label{exix}

\begin{figure}
\begin{center}
\includegraphics*[height=6cm]{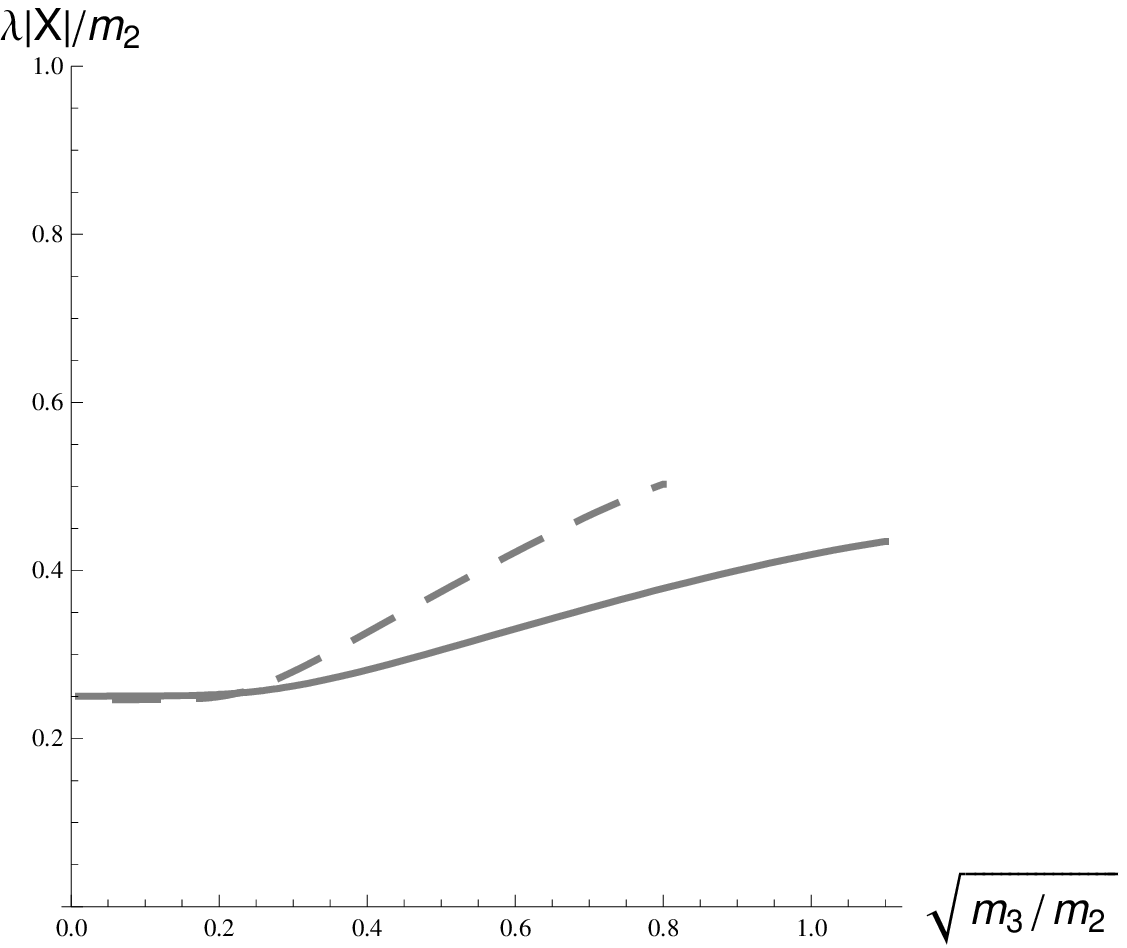}
\hspace{0.5cm}
\includegraphics*[height=6cm]{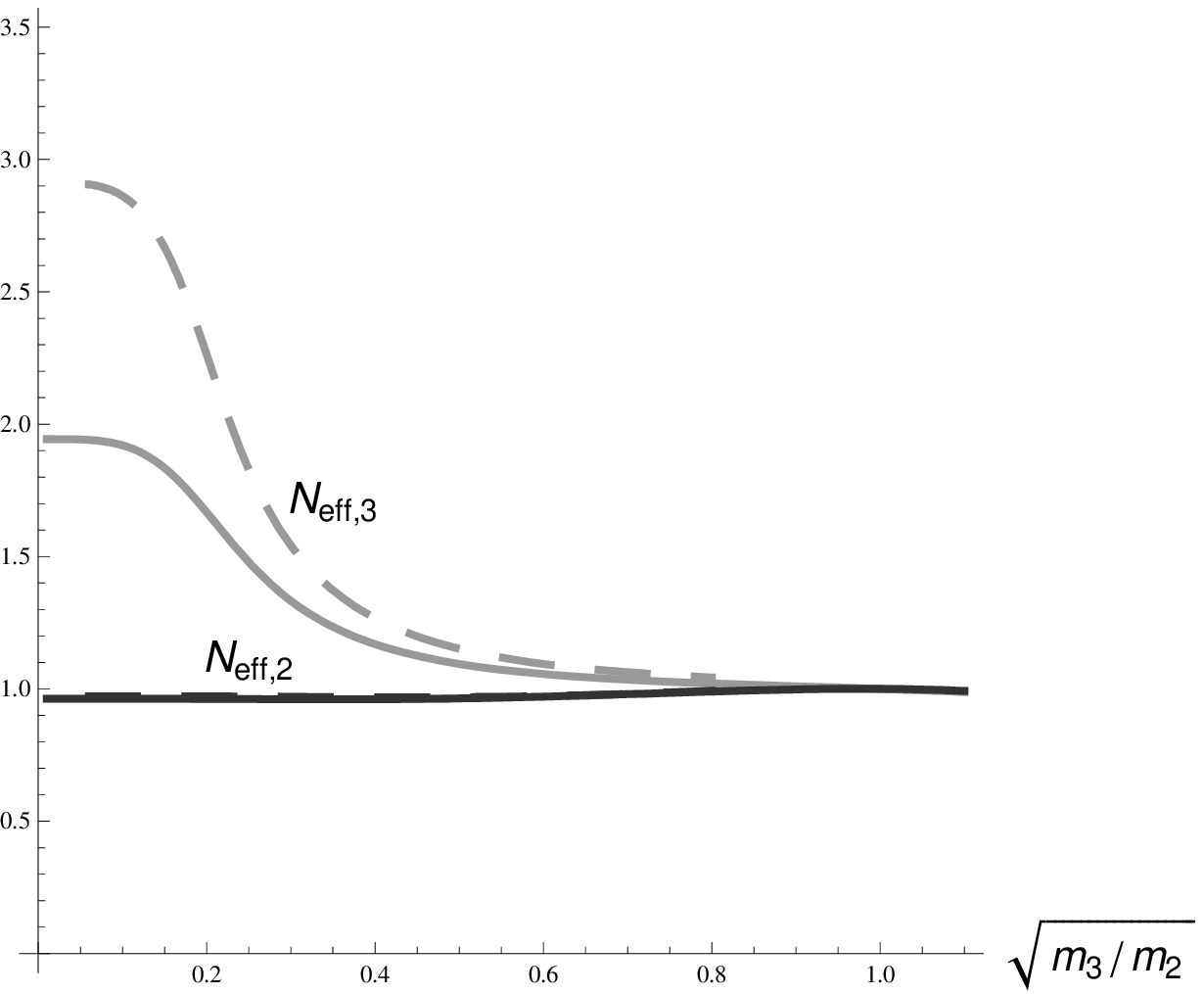}
\end{center}
\caption{\em The location $X$ of the supersymmetry breaking minimum 
and the effective number of messengers in the doublet and
triplet sectors, $N_{\mathrm{eff},2}$ and $N_{\mathrm{eff},3}$,
as functions of the ratio of the mass parameters in the doublet
and triplet sectors, $m_2$ and $m_3$,
for the models considered in Section \ref{exix} with $N=4$ (solid lines) and
$N=6$ (dashed lines).
\label{fdt0}}
\end{figure}

In this and the following section, we shall construct models with the messenger
and/or the O'Raifeartaigh sectors breaking the unified $SU(5)$. We assume that
it is more natural for the mass parameters than for the dimensionless couplings 
to break the unified gauge symmetry, as the former can arise from the
vevs of the Standard Model singlets in nontrivial $SU(5)$ representations. A well
known example of this idea is found in various realization of the doublet-triplet
splitting in the electroweak higgs sector of the unified theories,
which amounts to
giving large masses to the triplets which potentially mediate proton
decay while leaving the electroweak higgs doublets massless at this stage.

The simple model of Example (v)
in Appendix B is equipped with messenger parity for $\lambda_1=\lambda_2$,
but allows for different mass parameters for the doublet and triplet
components of the messenger multiplets. However, we find that with
$N=2$ pairs of messengers, as well as in the obvious $N=3$ generalization
of this model, the difference between
$N_{\mathrm{eff},2}$ and $N_{\mathrm{eff},3}$ is negligible
for ratios of the mass parameters for doublets and triplets ranging from $1/30$ to $30$,
so the doublet-triplet splitting in the messenger sector is not transmitted to the visible sector.
Hence the simplest model employs $N=4$ pairs of messengers with 
the mass matrices for doublets and triplets are, respectively,
\beq
\Mc^{(\mathbf{2})} = \left( \begin{array}{cccc} \lambda X & m_2 & 0 & 0 \\ 0 & \lambda X & m_2 & 0 \\ 0 & 0 & \lambda X & m_2 \\ 0 & 0 & 0 & \lambda X \end{array}\right)\, , \qquad
\Mc^{(\mathbf{3})} = \left( \begin{array}{cccc} \lambda X & m_2 & 0 & 0 \\ 0 & \lambda X & m_3 & 0 \\ 0 & 0 & \lambda X & m_2 \\ 0 & 0 & 0 & \lambda X \end{array}\right) \, .
\label{ex9dt}
\eeq
The value of $X$ at the minimum of the effective
potential and the effective messenger numbers are shown
in Figure \ref{fdt0}. 
The values of the effective messenger numbers can be understood in the following way.
For $m_3\ll m_2$, the mass matrix of the triplets is effectively split into two $2\times 2$
matrices studied in Example (v), while the mass matrix of the doublets remains irreducible.
As found in \cite{EOGM}, the mass matrix of the form of $\Mc^{(2)}$ ($\Mc^{(3)}$ with $m_3\to0$)
provides a supersymmetry breaking metastable minimum at $\lambda|X|/m_2=0.45$ ($0.25$),
and our numerical result interpolates between these two values. It also follows from
\cite{EOGM} that for $\lambda|X|/m_2<1$ the effective messenger number is approximately equal to
the number of irreducible sectors of the mass matrix; hence for $m_3\ll m_2$ we have
$N_{\mathrm{eff},2}\sim 1$ and $N_{\mathrm{eff},3}\sim 2$, whereas
for $m_3\sim m_2$ we obtain
$N_{\mathrm{eff},2}\approx N_{\mathrm{eff},3}\sim 1$.

By increasing the number of messengers to $N=6$, we can engineer a larger splitting
between
$N_{\mathrm{eff},2}$ and $N_{\mathrm{eff},3}$. Indeed, with the mass matrices:
\begin{displaymath}
\Mc^{(\mathbf{2,3})} = \left( \begin{array}{cccccc} \lambda X & m_2 & 0 & 0 & 0 & 0\\   0 & \lambda X & m_{2,3} & 0 & 0 & 0 \\  0 & 0 & \lambda X & m_2 & 0 & 0 \\ 0 & 0 & 0 & \lambda X & m_{2,3} & 0 \\ 0 & 0 & 0 & 0 & \lambda X & m_2 \\ 0 & 0 & 0 & 0 & 0 & \lambda X \end{array}\right)
\end{displaymath}
we find solutions with
$N_{\mathrm{eff},2}\sim 1$ and $N_{\mathrm{eff},3}\sim 3$;
the results are also shown in Figure \ref{fdt0}.

\subsection{Models with $N_{\mathrm{eff},2}<N_{\mathrm{eff},3}\approx1$}
\label{exx}

\begin{figure}
\begin{center}
\includegraphics*[height=7cm]{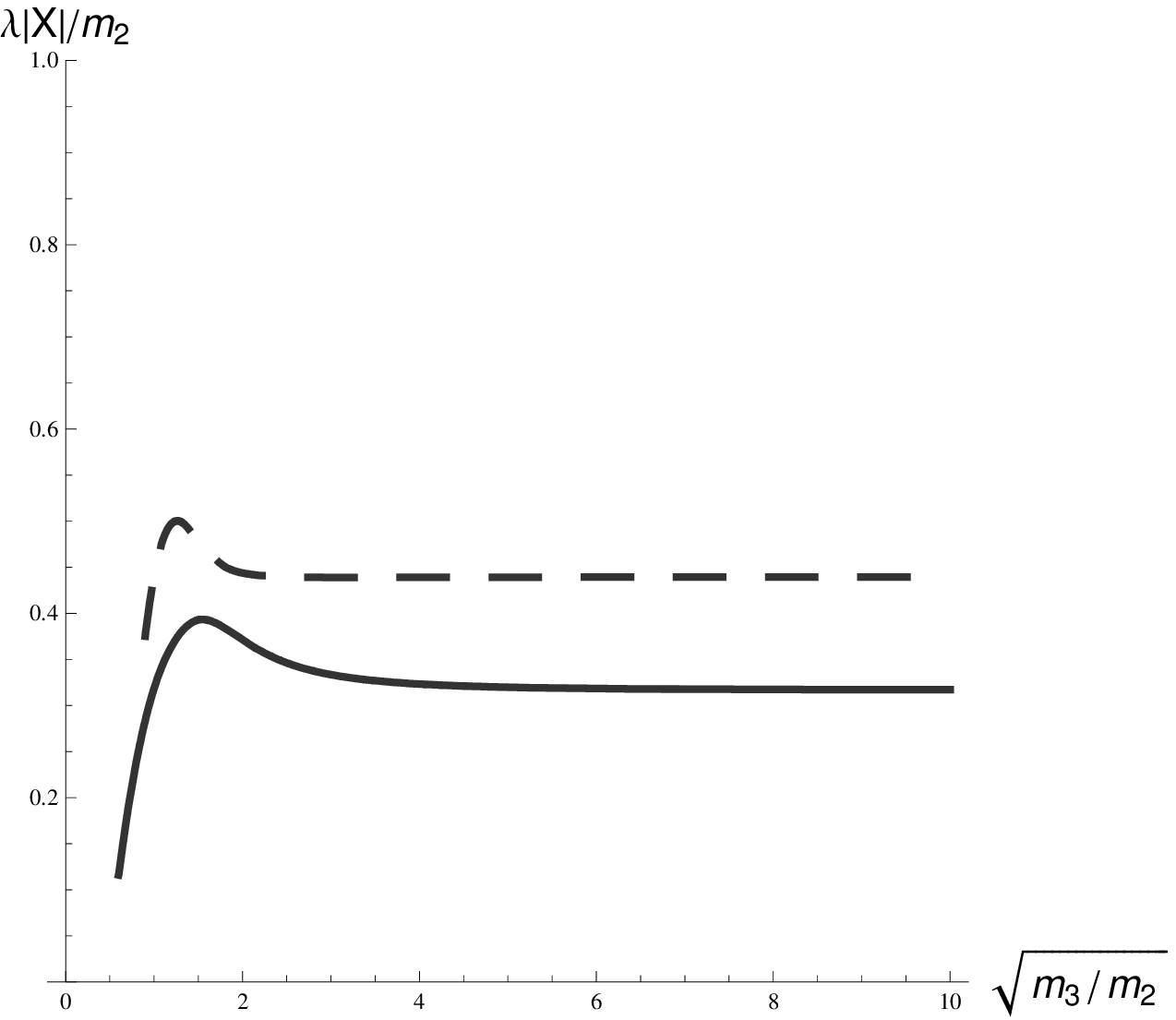}
\hspace{0.5cm}
\includegraphics*[height=7cm]{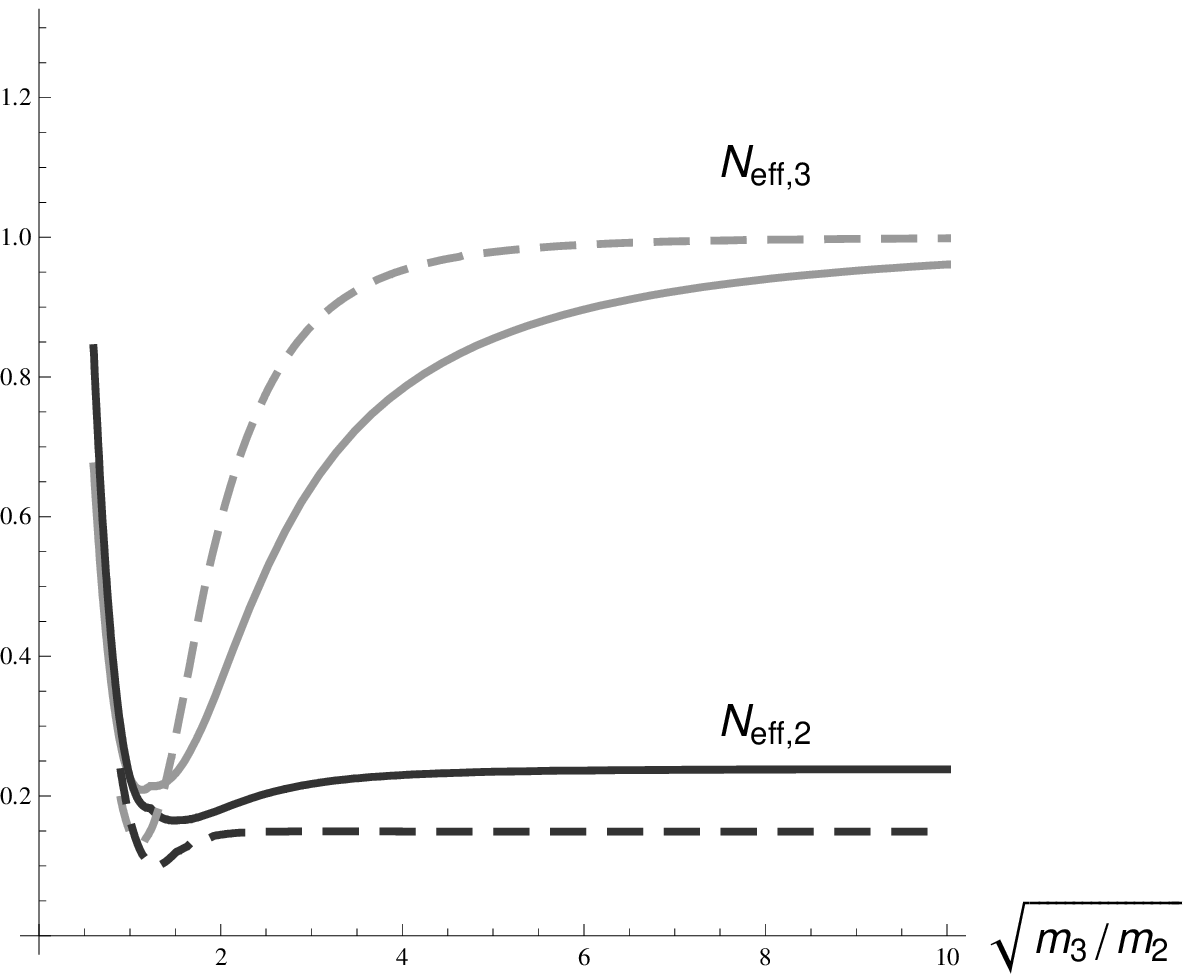}
\end{center}
\caption{\em The location $X$ of the supersymmetry breaking minimum 
and the effective number of messengers in the doublet and
triplet sectors, $N_{\mathrm{eff},2}$ and $N_{\mathrm{eff},3}$,
as functions of the ratio of the couplings in the doublet
and triplet sectors, $\lambda_2$ and $\lambda_3$,
for the models considered in Section \ref{exx}. Solid and dashed
lines correspond to models described by the interactions
(\ref{wex10}) and (\ref{wex10a}), respectively. 
\label{fdt5}}
\end{figure}

The model described in Example (iii) in Appendix B with $\tilde\rho=1$
is invariant under the messenger parity acting as:
\beq
\phi_{1,2,3} \to \tilde\phi_{3,2,1}^\ast \, , \tilde\phi_{1,2,3} \to \phi_{3,2,1}^\ast \, .
\eeq
We can therefore safely assume that the fields $\tilde\phi_i$ and $\phi_j$
belong to nontrivial representations of the Standard Model gauge
groups and they can act as messengers of the supersymmetry breaking
without obeying the unified $SU(5)$ symmetry.
Recall that we still need messenger fields which become massless in the
limit $X\to0$ in order to generate the gaugino masses at one loop.
Hence the simplest model built along these lines employs 
the following mass matrices for doublets and triplets:
\beq
\Mc^{(\mathbf{2})} = \left( \begin{array}{cccc} \lambda' X & 0 & 0 & 0 \\ 0  & m'_2 & \lambda X & 0 \\ 0 & 0 & m_2 & \lambda X \\ 0 & 0 & 0 & m'_2 \end{array}\right)\, , \qquad
\Mc^{(\mathbf{3})} = \left( \begin{array}{cccc} \lambda' X & 0 & 0 & 0 \\ 0  & m'_3 & \lambda X & 0 \\ 0 & 0 & m_3 & \lambda X \\ 0 & 0 & 0 & m'_3 \end{array}\right)\, .
\label{wex10}
\eeq
To ensure the gauge coupling unification, we assume that the determinants of $\Mc^{(\mathbf{2})}$
and  $\Mc^{(\mathbf{3})}$ are equal, which implies $(m'_2)^2m_2=(m'_3)^2m_3$. Interactions
 of the form
(\ref{wex10}) stabilize $X$ at a nonzero minimum for sufficiently large hierarchies of the mass 
parameters \cite{Shih} and we take $m_2=10m'_2$. Since the overall mass scale can be factored
out in (\ref{wex10}), we are then left with only one mass ratio $m_3/m_2$ 
on which the properties of the model depend.
We take $\lambda'=\lambda/1000$ to allow for a separation between
the value of $\langle X\rangle$ and the mass of the lightest messenger
(though the results for $N_{\mathrm{eff},r}$ would not change if $\lambda'=1/10)$.
In Figure \ref{fdt5}, we plot the position of the minimum as well as
the effective numbers of the messengers $N_{\mathrm{eff},2}$
and $N_{\mathrm{eff},3}$ in this model. We see that the asymptotic values
of the messenger numbers are $0.24$ and $1$.
We also see that $N_{\mathrm{eff},2}$ does not reach its asymptotic
value for $\lambda |X|/m_2 \to\infty$ (calculated in \cite{EOGM} and 
shown in Figure \ref{fneff} in Appendix C) since the position of the minimum corresponds
to $\lambda |X|/m_2 \sim 1/3$.
By increasing the number of
messengers, with explicit mass parameters, we can further change the ratio of 
$N_{\mathrm{eff},3}/N_{\mathrm{eff},2}$. For example, adding another
messenger pair so that the messenger interactions are now given by:
\beq
\Mc^{(\mathbf{2,3})} = \left( \begin{array}{ccccc} \lambda' X & 0 & 0 & 0 & 0\\ 0  & m'_{2,3} & \lambda X & 0 & 0 \\ 0 & 0 & m_{2,3} & \lambda X & 0 \\ 0 & 0 & 0 & m_{2,3} & \lambda X  \\ 0 & 0 & 0 & 0 & m'_{2,3} \end{array}\right)\, ,
\label{wex10a}
\eeq
again with $\mathrm{det}\Mc^{(\mathbf{2})}=\mathrm{det}\Mc^{(\mathbf{3})}$ and $m_2=10m'_2$,
leads to results shown in Figure \ref{fdt5}
with asymptotic values of the effective messenger numbers equal $0.15$
and $1$. By increasing the number of messengers with masses proportional to $\langle X\rangle$,
we can increase all $N_{\mathrm{eff},r}$, e.g.~adding one pair of such messengers to the
model defined by (\ref{wex10a}) gives $N_{\mathrm{eff},2}\approx 0.7$ and 
$N_{\mathrm{eff},3}\approx 2$.

All models have a stable supersymmetric vacuum.
Phenomenologically consistent models with broken $SU(5)$ symmetry can be constructed,
with different effective doublet and triplet messenger numbers. 
With sufficiently large mass or coupling hierarchies between the doublet and
triplet sectors, these effective messenger numbers depend only on
the $R$-charges of the messenger fields, rather than on their masses
and couplings.
Embedding these models in supergravity still brings about
an upper bound on $X$ discussed in Section \ref{sec3}, though
due to a large number of fields over which the trace runs in the effective
K\"ahler potential (\ref{eff2}), one-loop corrections compete more efficiently
with the supergravity corrections. For example, in the model given by (\ref{wex10a}) 
we obtain $X\simlt 10^{-2}M_P$.

\section{Physical implications}
\label{sec5}

Gravity coupled to the messenger sector via ORT supersymmetry breaking sector
raises the question about the role of gravity mediation versus gauge mediation of
supersymmetry breaking in generating the soft supersymmetry breaking mass terms in the 
MSSM.  The gravity mediation contribution to the scalar  masses is of the order of the
gravitino mass
$
m_{3/2}=F_X/(\sqrt{3} M_P)
$,
where $F_X=F$.  
The gauge mediated contribution to
the soft scalar and gaugino masses are (up to the (effective) messenger
number and $O(1)$ coefficients) given by the scale:
\begin{equation}
\label{softabove}
m_\mathrm{gaugino} = \frac{\alpha}{4\pi} \frac{\lambda_\mathrm{mess} F_X}{\Mc_\mathrm{mess}} \, ,
\end{equation}
where $\Mc_\mathrm{mess}$ is the average mass of those messengers which
give significant contributions to gaugino masses, 
the effective messenger coupling
$\lambda_\mathrm{mess}$ is defined as $\Mc_\mathrm{mess}/\langle X\rangle$, 
$\lambda_\mathrm{mess} F_X$
is the mass squared splitting in the messenger supermultiplets
and $\alpha$ is an appropriate coupling constant.
With the simplest possible messenger sector consisting of one pair
of fields in the $\mathbf{5}+\mathbf{\bar{5}}$ representations of
the unified gauge symmetry group $SU(5)$, we identify 
$\lambda_\mathrm{mess}=\tilde \lambda$, but we would like to keep our
discussion general enough to include all the situations discussed
in the previous Sections, in particular, models with large mass splittings
within the messenger sector.
For example, in models discussed in Section \ref{exx} 
$\lambda_\mathrm{mess}$ can be identified with $\lambda'$,
ranging from $10^{-3}$ to $10^{-1}$, even when all other universal
couplings $\lambda$. are close to 1.
The main result of Section \ref{sec3} is that in the simple ORT models 
coupled to gravity there
exists an upper bound on the value of $X$ of order
of $10^{-3}M_P$ in the metastable supersymmetry breaking minimum.  
An immediate conclusion is then that in metastable vacua
\begin{equation}
\frac{m_{3/2}}{m_\mathrm{gaugino}} = \frac{4\pi}{\alpha\sqrt{3}} \frac{\langle X\rangle}{M_P} < \mathrm{O}(10^{-1}) \, ,
\label{justabove}
\end{equation}
where in the last step we used conservative values of $\alpha=0.04$ and
$\langle X\rangle=10^{-3}M_P$.
Given the conservative nature of these numbers, we infer that
in gauge mediation models with ORT sector responsible for $F$-term supersymmetry breaking
coupled to gravity, in metastable vacua gravity mediation contribution to squark mass squared  is naturally suppressed to 
at least $\mathrm{O}(1\%)$ level, almost
sufficient for avoiding FCNC problem in the squark sector \cite{Nir}.
The tunneling rate from the metastable vacuum at $X\ll M_P$ 
towards the tree-level Polonyi vacuum at $X\sim M_P$ is very small, and, 
in the order-of-magnitude approximation, see \cite{Lee:1985uv}, is given by
\begin{equation}
\Gamma = \frac{1}{\tau_D} \approx 0.1 (\bar m/M_P)^4 M_P e^{-30 \frac{\bar m^3 M_P}{F^2}} \, , 
\end{equation}
where $\bar m$ is the average mass in the O'Raifeartaigh sector. 
This corresponds to a lifetime $\tau_D$ much longer than the age of 
the Universe.

From eq.\ (\ref{justabove}) we conclude that for 
$m_\mathrm{gaugino}\approx  100(1000) \mathrm{GeV})$ the gravitino
mass $m_{3/2} < \mathrm{O}(10,100\,\mathrm{GeV})$. 
(Similar values for the gravitino mass have been
discussed in \cite{Nomura,Nir}; 
here we show that this is a generic upper bound for
gravitino mass in gauge mediation models coupled to gravity.)

The Giudice-Masiero mechanism of generating the $\mu$-term in the effective low-energy superpotential relies on the presence in the high-energy 
K\"ahler potential of an interaction term of the form 
\begin{equation}
\delta K = \frac{1}{2} \frac{X^\dagger}{M_P} H_u H_d \, + \, {\rm h.c.} \, .
\end{equation}
With the help of the well known formulae, see \cite{Kaplunovsky:1993rd}, one obtains this way 
\begin{equation}
|\mu| = \left|m_{3/2} \frac{X^\dagger}{M_P} - \frac{F^{\bar{X}}}{M_P}\right|  = m_{3/2} \left|\frac{X^\dagger}{M_P} - \sqrt{3} \right| \, ,
\label{gmm1}
\end{equation}
where the cancellation of the cosmological constant is assumed, and 
\begin{equation}
B= \pm m_{3/2} \, .
\label{gmm2}
\end{equation}
Hence, if one chooses the gravitino mass to be of the order of $100$ GeV, one finds $\mu$ and $B$ of this order of magnitude, which, as we shall see 
somewhat later, may be consistent with radiative breaking of the 
electroweak symmetry. 

We now proceed to a more detailed discussion
of  the  simple but viable ORT models
with and without the doublet-triplet splitting violating the
$SU(5)$ mass relations, studied in Section \ref{sec2} and \ref{sec4}. 
We shall be specially interested in phenomenology of the models with
heavy messengers, i.e. with the values of $X$ close to their upper bound.

In Table \ref{tab1}  we collect a sample of models discussed in the previous
sections; these models are characterized by the effective number of messengers.
Cases A-C correspond to the usual models of gauge mediation
with equal number of messengers in the doublet and triplet sectors.
cases D and E are described in Section \ref{exix}, and cases F and G in
Section \ref{exx}.

\begin{table}
\begin{center}
\begin{tabular}{c|ccccccc}
 & A & B & C & D & E & F & G \\
\hline
$N_{\mathrm{eff},2}$ & 1 & 4 & 7 & 1 & 1 & 0.24 & 0.15 \\
$N_{\mathrm{eff},3}$ & 1 & 4 & 7 & 2 & 3 & 1 & 1 
\end{tabular}
\end{center}
\caption{\em Parametrization of some interesting spectra of
conventional models of gauge mediation ($N_{\mathrm{eff},2}=N_{\mathrm{eff},3}$) and of models described in Sections \ref{exix} and \ref{exx}. \label{tab1}}
\end{table}

\begin{figure}
\begin{center}
\includegraphics*[height=6cm]{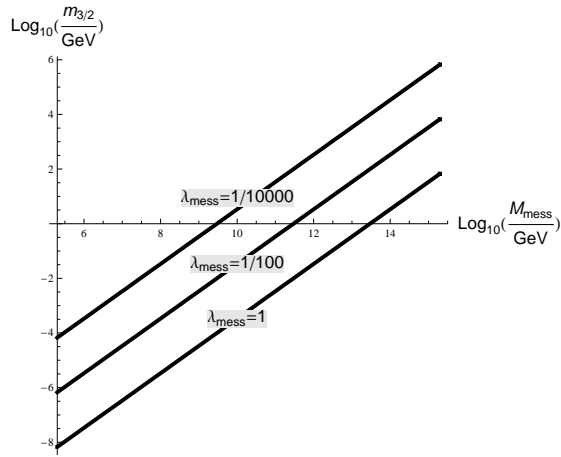}
\end{center}
\caption{\em The relation (\ref{przel}) between $\Mc_\mathrm{mess}$
and $m_{3/2}$ for different values of $\lambda_\mathrm{mess}$
and for $F\,\partial_X\ln\mathrm{det}\,\Mc=140\,\mathrm{TeV}$.
\label{fprzel}}
\end{figure}

With fixed $N_{\mathrm{eff},2}$ and $N_{\mathrm{eff},3}$,
the effective continuous parameters of these models of gauge mediation
can be chosen, e.g., as the messenger mass scale $\Mc_\mathrm{mess}$
and the gaugino (say, gluino) initial mass.
We impose the 
boundary conditions for the soft masses
(\ref{mass:gaugino}) and (\ref{mass:scalar})
at the scale $\Mc_\mathrm{mess}$.
The two parameters are related to the gravitino mass by (\ref{justabove})
or, more explicitly:
\beq
m_{3/2} = \frac{4\pi}{\sqrt{3}\alpha} \,\frac{\Mc_\mathrm{mess}}{M_P} \,\frac{m_\mathrm{gaugino}}{\lambda_\mathrm{mess}}
\label{przel}
\eeq
In the following, we take $F\,\partial_X\ln\mathrm{det}\,\Mc=140\,\mathrm{TeV}$
in (\ref{mass:gaugino}), which approximately corresponds to the gluino mass
of $1\,\mathrm{TeV}$ at the electroweak scale,
and vary the
messenger mass between $\Mc_\mathrm{mess}=2\cdot10^5\,\mathrm{GeV}$
and $\Mc_\mathrm{mess}=2\cdot10^{15}\,\mathrm{GeV}$. The lower limit
comes from the requirement that $F/\bar m^2<1$, for which the effective
K\"ahler potential technique applies, the upper limit follows from
our results in Section \ref{sec3}. For these values of the parameters, we plot
in Figure \ref{fprzel}
$m_{3/2}$ as a function of $\Mc_\mathrm{mess}$ for different values
of $\lambda_\mathrm{mess}$.

\begin{figure}
\begin{center}
\includegraphics*[height=7cm]{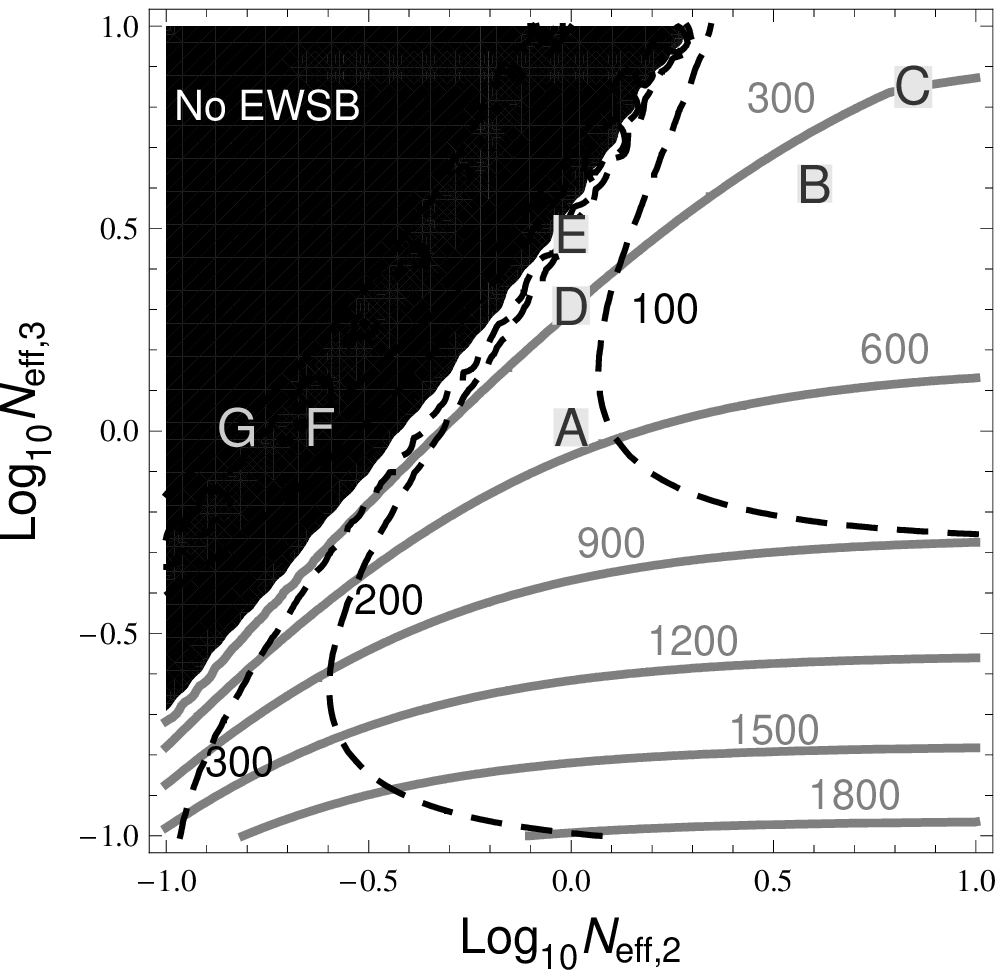}
\hspace{0.5cm}
\includegraphics*[height=7cm]{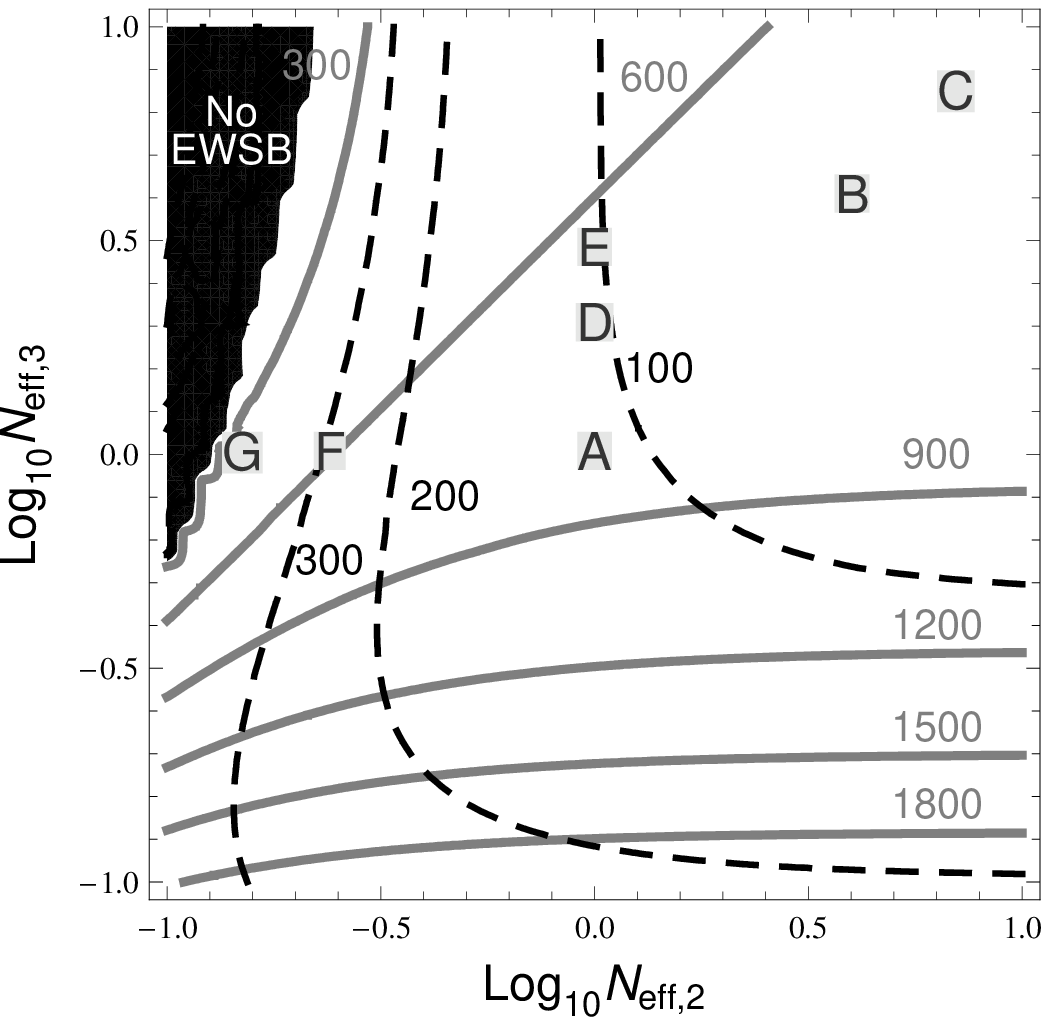}
\end{center}
\caption{\em Values of $\mu$ (solid gray lines) and $B$ (dashed black) 
in $\mathrm{GeV}$ as functions of the effective numbers of messengers
$N_{\mathrm{eff},2}$ and $N_{\mathrm{eff},3}$ for two messenger scales
$\Mc_\mathrm{mess}=2\cdot10^5\,\mathrm{GeV}$ (left panel) and
$\Mc_\mathrm{mess}=2\cdot10^{15}\,\mathrm{GeV}$ (right panel).
Black regions correspond to the absence of electroweak symmetry breaking.
Points corresponding to spectra A-G defined in Table \ref{tab1}
are denoted by appropriate letters. 
\label{fewsb}}
\end{figure}

The requirement of proper electroweak symmetry breaking fixes the MSSM parameters
$\mu$ and $B$ in terms of the initial  values of soft masses.
In Figure \ref{fewsb} we show the predictions for $\mu$
(solid gray lines, values in GeV) and for $B$ (dashed black lines,
values in GeV) as function of $N_{\mathrm{eff},2}$ and $N_{\mathrm{eff},2}$
for the two extremal values of the messenger masses, 
 $\Mc_\mathrm{mess}=2\cdot10^5\,\mathrm{GeV}$ (left)
and $\Mc_\mathrm{mess}=2\cdot10^{15}\,\mathrm{GeV}$ (right).
These values of $\mu$ and $B$ are given at the scale $\Mc_\mathrm{mess}$.
The values of $N_{eff}$ for the seven cases listed in Table 1 are denoted by the corresponding letters.
We chose $\tan\beta=10$, different choices of this parameter do not
affect $\mu$ much, while $B$ is very sensitive to the value of $\tan\beta$
(this dependence is quantitatively discussed in Appendix D).
The results shown in Figure \ref{fewsb} are obtained from numerical calculation based on 1(2)-loop
RG equations for the dimension(less) parameters of the MSSM
(with the exception of the $Y=+1/2$ higgs mass parameter, for which
the 2-loop RGE was used).
As we show in Appendix D, 
they can be understood quite easily, e.g., 
with help of the analytical
solutions to the RGE for nonuniversal soft masses given 
in \cite{bottom-up}.
We note from Figure \ref{fewsb} that for the same values of $N_{\mathrm{eff},2}$ and $N_{\mathrm{eff},3}$ the ratio $\mu/B$ increases
with the messenger mass $\Mc_{mess}$. Furthermore, for $N_{\mathrm{eff},2}\approx N_{\mathrm{eff},3}$
and for $N_{\mathrm{eff},2}>N_{\mathrm{eff},3}$, $\mu/B \gg1$  whereas for $N_{\mathrm{eff},2}< N_{\mathrm{eff},3}$ we have
$\mu/B \approx 1$.  This is interesting in case of heavy messengers (right panel in Figure \ref{fewsb}).
As we discuss in some detail in Appendix D,
there exist then solutions with $\mu \sim B \approx m_{3/2} \approx O(100)\, \mathrm{GeV}$,
opening the possibility of generating $\mu$ and $B$ by the  Giudice-Masiero mechanism.

\begin{figure}
\begin{center}
\includegraphics*[height=4.5cm]{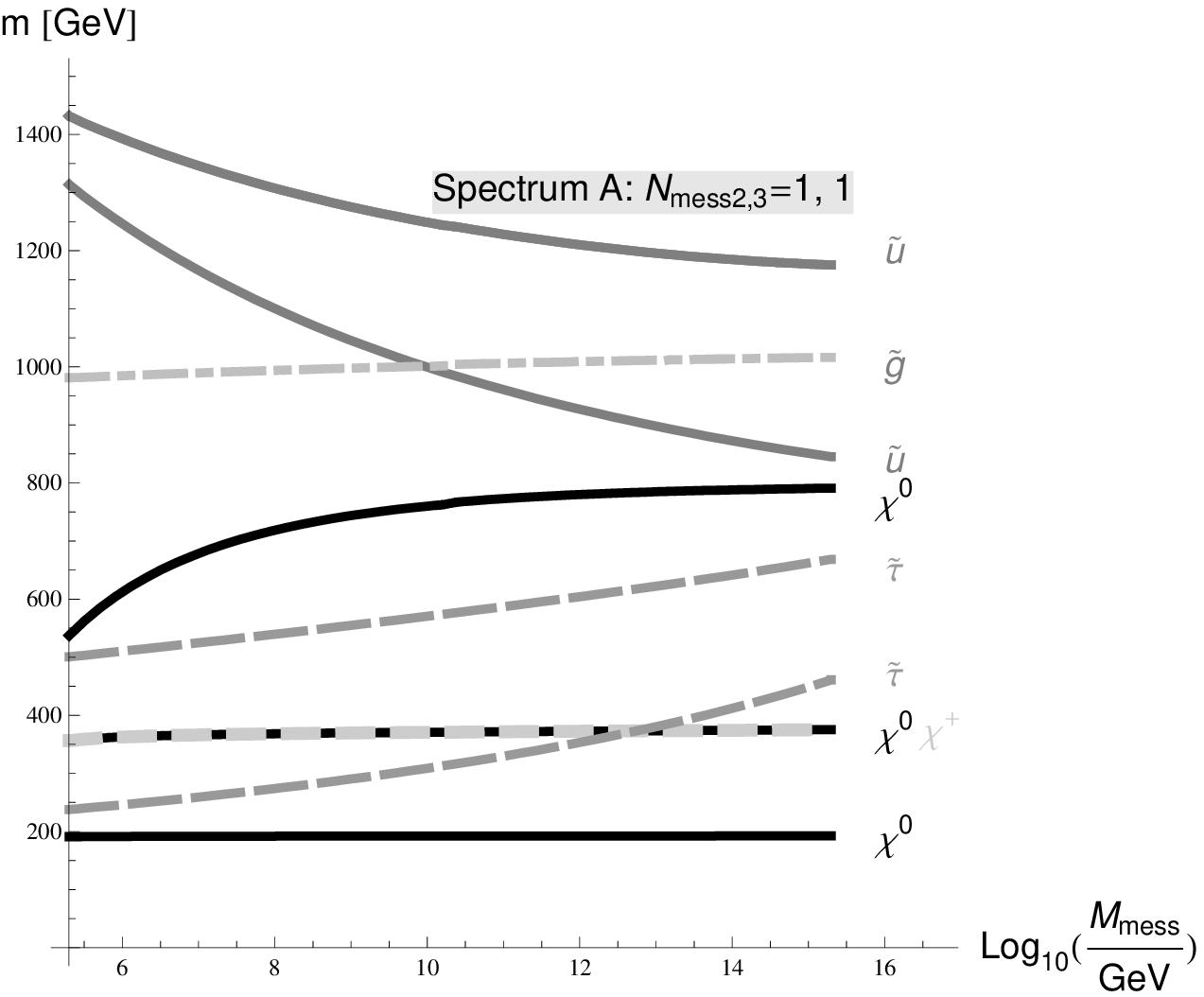}
\hspace{0.0cm}
\includegraphics*[height=4.5cm]{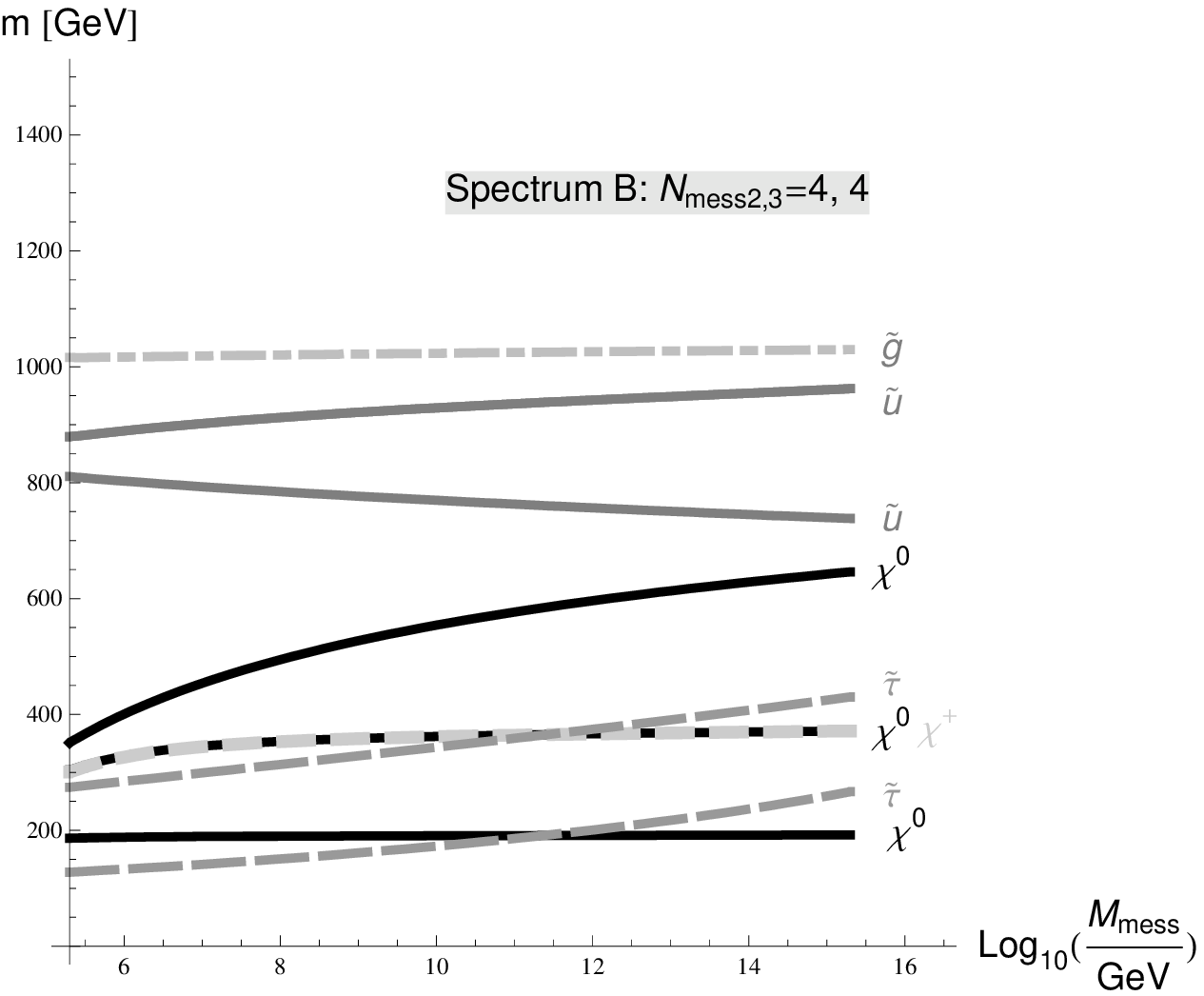}
\hspace{0.0cm}
\includegraphics*[height=4.5cm]{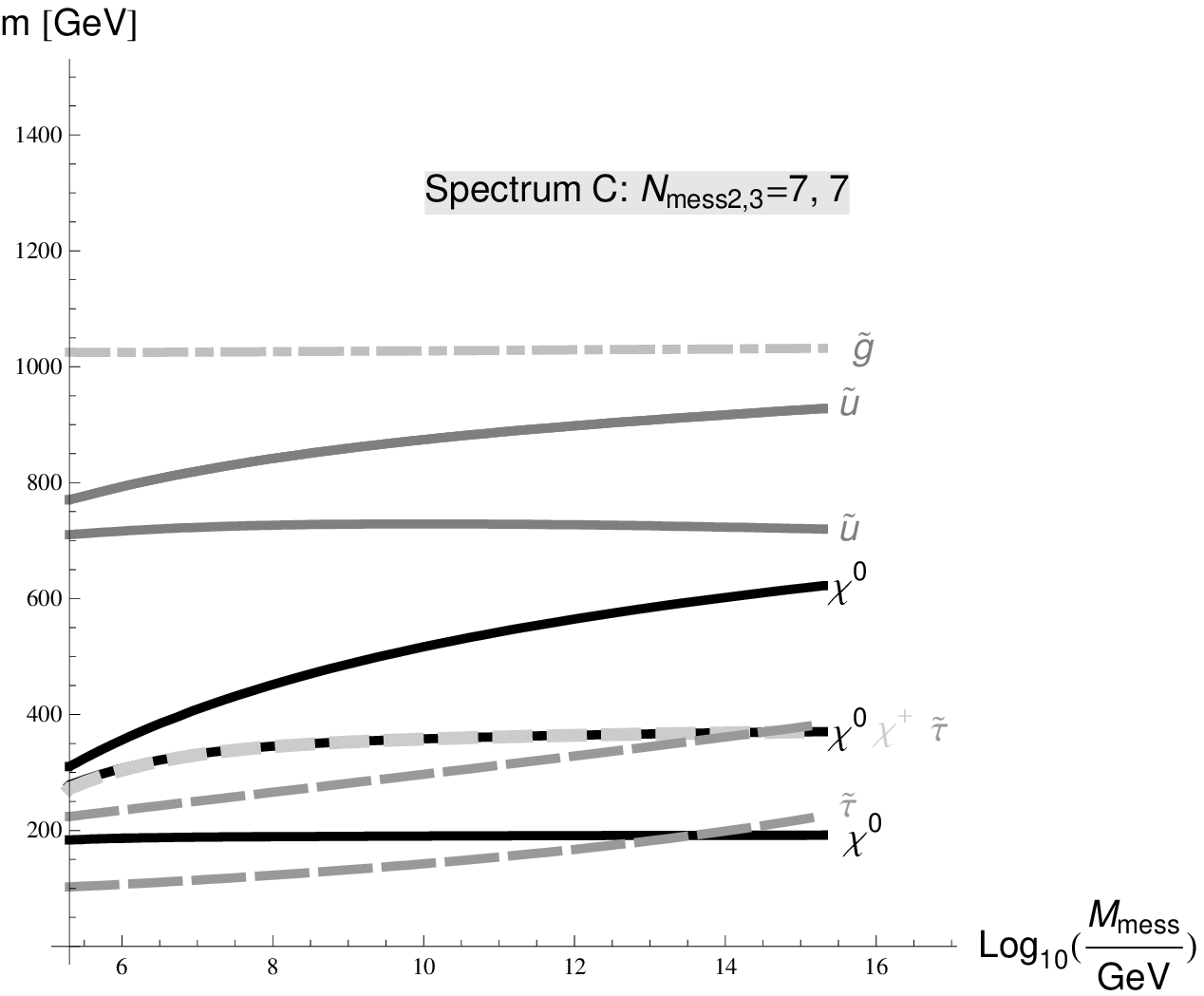}
\end{center}
\caption{\em Masses of the three lightest neutralinos, the lightest
chargino, the lightest left- and right-handed sleptons, the two lightest
stops and the gluino as functions of the messenger scale
for spectra A-C defined in Table \ref{tab1}.
\label{fspr1}}
\end{figure}

\begin{figure}
\begin{center}
\includegraphics*[height=6cm]{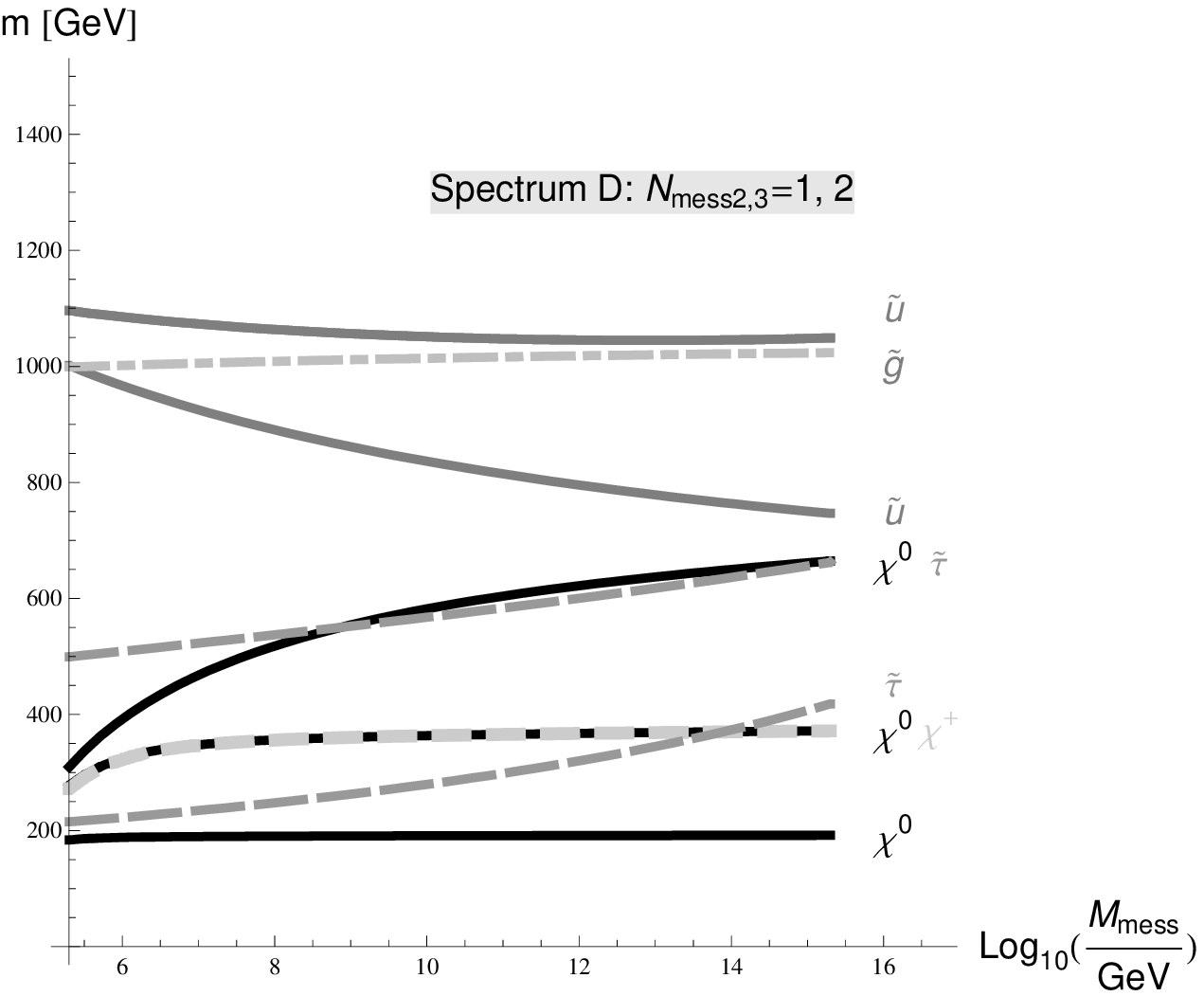}
\hspace{0.5cm}
\includegraphics*[height=6cm]{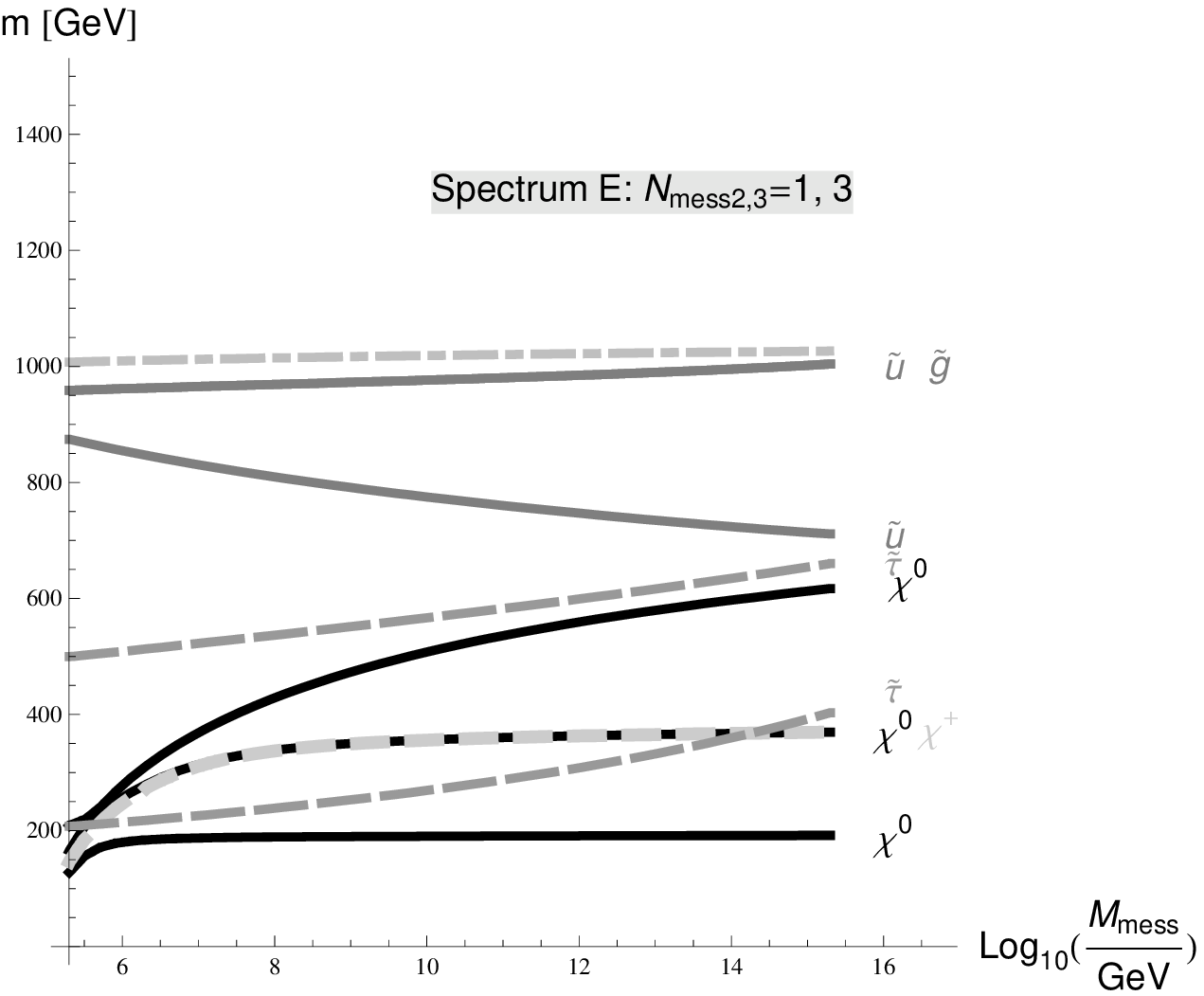}
\end{center}
\caption{\em  Masses of the three lightest neutralinos, the lightest
chargino, the lightest left- and right-handed sleptons, the two lightest
stops and the gluino as functions of the messenger scale
for spectra D and E defined in Table \ref{tab1}.
\label{fspr2}}
\end{figure}

\begin{figure}
\begin{center}
\includegraphics*[height=6cm]{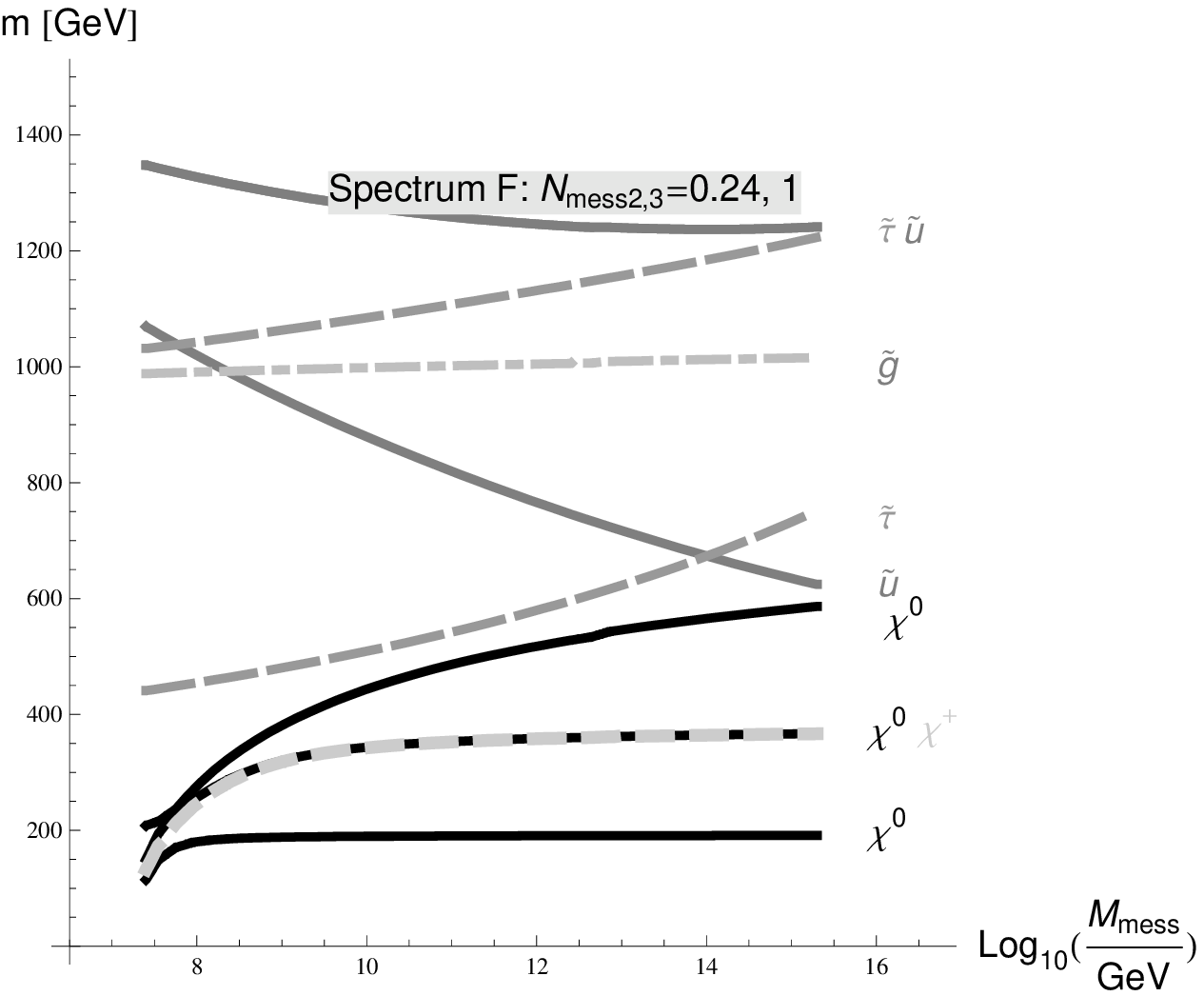}
\hspace{0.5cm}
\includegraphics*[height=6cm]{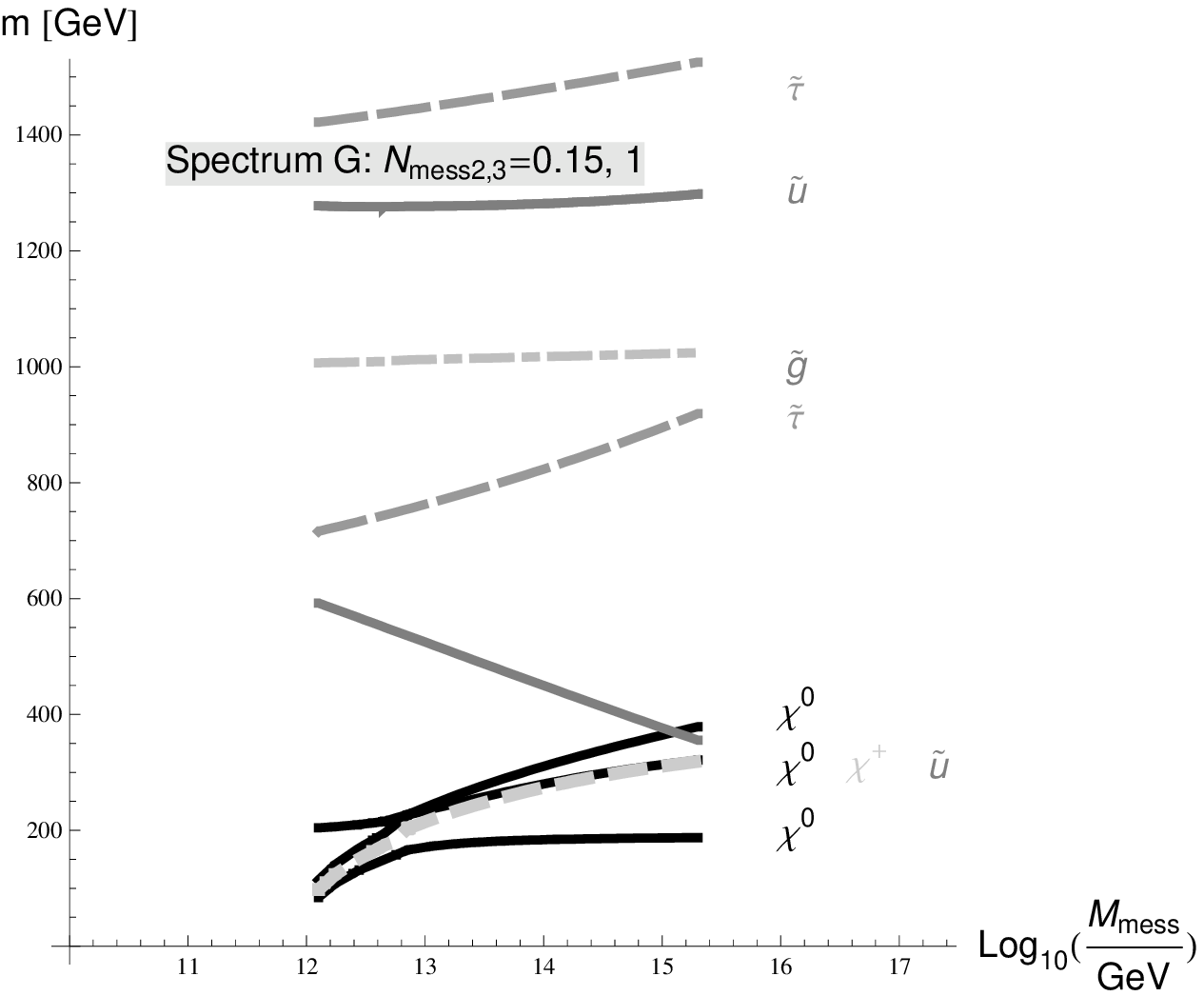}
\end{center}
\caption{\em  Masses of the three lightest neutralinos, the lightest
chargino, the lightest left- and right-handed sleptons, the two lightest
stops and the gluino as functions of the messenger scale
for spectra F and G defined in Table \ref{tab1}.
\label{fspr3}}
\end{figure}

Our results for the supersymmetric spectra are shown 
in Figures \ref{fspr1}-\ref{fspr3}. There we show the masses of
the three lightest neutralinos, the lightest
chargino, the lightest left- and right-handed sleptons, the two lightest
stops and the gluino.
The dependence on the messenger masses is rather 
weak, except  for cases F and G where a portion of the assumed
messenger mass range does not lead to a correct electroweak symmetry
breaking, as seen by comparing the two panels of Figure \ref{fewsb}.
Generically, the NLSP is always bino-like neutralino. There are, however, two interesting
exceptions. One is the stau as NLSP for large number of messengers (with $N_{eff,2}=N_{eff,3}$). This effect is present also for models with heavy messengers, with gravitino mass even
up to 100 GeV. The second exception is for models with $N_{eff,2} \ll N_{eff,3}$ where the lightest neutralino becomes higgsino-like if the scale 
$\Mc_\mathrm{mess}$ is such that various contributions to the mass of the 
$Y=+1/2$ higgs approximately cancel.
This regime with small $\mu$ and a large higgsino admixture in
the lightest neutralino corresponds,
in Figure \ref{fspr3}, to the situation in which
the lines denoting the neutralino masses 
exhibit level crossing.
As we show in Section \ref{exx} and in Appendix C, 
$\lambda_\mathrm{mess}$ can be as small as $10^{-3}$ (or even
smaller).
Thus, for gravitino masses in the range 10-100 GeV the higgsino-like neutralino as  NLSP
is also an open possibility.

\section{Summary and conclusions}
\label{sec6}

We have investigated O'Raifeartaigh-type models for $F$-term supersymmetry
breaking in gauge mediation scenarios in the presence of gravity.
In Section \ref{sec2}  the vacuum structure of a broad class of models has been studied in the global supersymmetry limit.
Gravity effects have been discussed in Section \ref{sec30}. 
The gravity sector may include properly stabilized
moduli. In Section \ref{sec4}, we have discussed models  with broken
$SU(5)$ symmetry in the messenger sector but equipped with messenger parity
(and the disastrous consquences of the lack thereof).

The main conclusion is that, after coupling to gravity, the vacuum structure of those
models is such that in metastable vacua gauge mediation is always dominant and gravity mediation contribution to scalar masses is
suppressed to the level below 1 percent, almost sufficient for avoiding FCNC problem.
Close to that limit, gravitino mass can be in the range 10-100 GeV, opening several interesting
possibilities for gauge mediation models, which are briefly discussed in Section 5.
One is that in models with broken $SU(5)$ symmetry, the values of $\mu$ and $B$ fixed by
requiring the electroweak symmetry breaking include the region $\mu \approx B \approx m_{3/2} \sim O(100)\, \mathrm{GeV}$, thus allowing for Giudice-Masiero mechanism for $\mu$
and $B\mu$ generation. We also show sparticle spectrum as a function of the gravitino mass.
The NLSP is generically bino-like neutralino but exceptionally can be stau or higgsino.

\section*{Acknowledgments}
\vspace*{.5cm}
\noindent This work was partially supported by the 
EC 6th Framework Programme MRTN-CT-2006-035863,  
by the EC 6th Framework Programme MRTN-CT-2004-503369, 
by the  grant MNiSW  N202 176 31/3844 and  by TOK Project  
MTKD-CT-2005-029466. 
KT is supported by the US Department of Energy.
ZL and KT thank APC Paris for hospitality.
KT is indebted to D.~Morrissey and A.~Pierce for discussions. 

\section*{Appendix A}

Here we outline the perturbative procedure of calculating the eigenvalues of
$ \Mc^\dagger \Mc$. 
In the limit $|X|\to0$, the eigenvalues are  
$\mu^{(k)}_0=m_k^2$ and the corresponding eigenvectors 
are ${}^0v^{(k)}_i=\delta_{ki}$. 
At the $n$-th level of the expansion in $X$, 
we find the coefficients of the $|X|^n$ contribution to the eigenvalues
$\mu^{(k)}_n$ and the eigenvectors ${}^nv^{(k)}_i$ from the perturbed
secular equation:
\bea
&m_i^2 \left( {}^nv^{(k)}_i\right) + \sum_{j=1}^N(m_j\lambda_{ji}^\ast\bar X+m_i\lambda_{ij}X) \left( {}^{n-1}v^{(k)}_j\right) + \sum_{j,\ell=1}^N \lambda_{\ell i}^\ast \lambda_{\ell j} \bar{X}X  \left( {}^{n-2}v^{(k)}_j\right) = & \nonumber \\ &=\sum_{\ell=0}^n \mu^{(k)}_\ell \left({}^{n-\ell}v^{(k)}_i\right)&
\eea
assuming $\mu^{(k)}_n={}^nv^{(k)}_i=0$ for $n<0$. 
The coefficients $\mu^{(k)}_n$ can be equivalently defined as:
\begin{equation}
\mu^{(k)}_{2j} = \left. \frac{1}{j!} \frac{\bar{m}^{2j}}{\bar{\lambda}^{2j}}\left(\frac{\partial^2}{\partial\bar X\partial X}\right)^j  \Mc_k^2 \right|_{|X|=0} \, ,
\end{equation}
where $ \Mc_k^2$ are the eigenvalues of 
$( \Mc^\mathrm{ns})^\dagger \Mc^\mathrm{ns}$, 
introduced in Section 2. 
The first functions $f_{2\ell}$ are given by:
\begin{eqnarray}
f_2 &=& \frac{1}{\bar m^2}\sum_{k=1}^N \left( \mu^{(k)}_2\ln\frac{m_k^2}{Q^2}+\mu^{(k)}_2\right) \\
f_4 &=& \frac{1}{\bar m^4}\sum_{k=1}^N \left( \mu^{(k)}_4\ln\frac{m_k^2}{\bar m^2}+\frac{\bar m^2}{2}\frac{(\mu^{(k)}_2)^2}{m_k^2}\right) \\
f_6 &=& \frac{1}{\bar m^6} \sum_{k=1}^N \left( \mu^{(k)}_6\ln\frac{m_k^2}{\bar m^2}+\bar m^2\frac{\mu^{(k)}_2\mu^{(k)}_4}{m_k^2}-\frac{\bar m^4}{6}\frac{(\mu^{(k)}_2)^3}{m_k^4}\right)
\end{eqnarray}
Using this procedure, we can express the coefficients $\mu^{(k)}_2$ 
and $\mu^{(k)}_4$ in terms of the original parameters of the model, 
the relevant general formulae are, however, rather lengthy.
They significantly simplify under assumption that
$R(\phi_1)>R(\phi_1)>\ldots>R(\phi_N)$
and $R(\tilde\phi_1)<R(\tilde\phi_2)<\ldots<R(\tilde\phi_N)$,
i.e.~the fields have different $R$ charges,
since the matrix of the couplings has then only a few nonzero entries,
$\lambda_{ij}=\bar\lambda q_i e^{\imath\varphi_i}\delta_{i+1,j}$.
We then obtain:
\begin{eqnarray}\frac{\mu^{(k)}_2}{\bar{m}^2} &=& \frac{\rho_k^2 q_{k-1}^2}{\rho_k^2-\rho_{k-1}^2} + \frac{\rho_k^2 q_k^2}{\rho_k^2-\rho_{k+1}^2} \\
\frac{\mu^{(k)}_4}{\bar{m}^2} &=& \frac{\rho_k^2\rho_{k-1}^2q_{k-1}^2q_{k-2}^2}{(\rho_k^2-\rho_{k-1}^2)^2(\rho_k^2-\rho_{k-2}^2)}+\frac{\rho_k^2\rho_{k+1}^2q_k^2q_{k+1}^2}{(\rho_k^2-\rho_{k+1}^2)^2(\rho_k^2-\rho_{k+2}^2)} - \nonumber\\
&& - \frac{\rho_k^2(\rho_k^4-\rho_{k-1}^2\rho_{k+1}^2)q_k^2q_{k-1}^2}{(\rho_k^2-\rho_{k-1}^2)^2(\rho_k^2-\rho_{k+1}^2)^2} - \frac{\rho_k^2\rho_{k-1}^2q_{k-1}^4}{(\rho_k^2-\rho_{k-1}^2)^3}- \frac{\rho_k^2\rho_{k+1}^2q_k^4}{(\rho_k^2-\rho_{k+1}^2)^3}
\\
\frac{\mu^{(k)}_6}{\bar{m}^2} &=& \frac{\rho_k^2\rho_{k-1}^2\rho_{k-2}^2q_{k-1}^2q_{k-2}^2 q_{k-3}^2}{(\rho_k^2-\rho_{k-1}^2)^2(\rho_k^2-\rho_{k-2}^2)^2(\rho_k^2-\rho_{k-3}^2)} + \frac{\rho_k^4\rho_{k-1}^2q_{k-1}^2q_{k-2}^4}{(\rho_k^2-\rho_{k-1}^2)^3(\rho_k^2-\rho_{k-2}^2)^2} + \nonumber\\
&& + \frac{\rho_k^2\rho_{k-1}^2(-2\rho_k^4+2\rho_{k-1}^2\rho_{k+1}^2+\rho_k^2\rho_{k-2}^2-\rho_k^2\rho_{k-1}^2)q_{k-1}^4q_{k-2}^2}{(\rho_k^2-\rho_{k-1}^2)^4(\rho_k^2-\rho_{k-2}^2)^2}+ \frac{\rho_k^2\rho_{k-1}^2(\rho_k^2+\rho_{k_1}^2)q_{k-1}^6}{(\rho_k^2-\rho_{k-1}^2)^5} - \nonumber\\
&& -\frac{\rho_k^2\rho_{k-1}^2(3\rho_k^6+\rho_k^2\rho_{k-2}^2\rho_{k+1}^2+\rho_{k-1}^2\rho_{k-2}^2\rho_{k+1}^2-\rho_k^4(\rho_{k-1}^2+2\rho_{k-2}^2+2\rho_{k+1}^2))q_kq_{k-1}^2q_{k-2}^2}{(\rho_k^2-\rho_{k-1}^2)^3(\rho_k^2-\rho_{k-2}^2)^2(\rho_{k}^2-\rho_{k+1}^2)} +\nonumber \\
&& +\frac{\rho_k^2(\rho_k^8+2\rho_k^6\rho_{k-1}^2-6\rho_k^4\rho_{k-1}^2\rho_{k+1}^2+2\rho_k^2\rho_{k-1}^2\rho_{k+1}^4+\rho_{k-1}^4\rho_{k+1}^4)q_{k-1}^4q_k^2}{(\rho_k^2-\rho_{k-1}^2)^4(\rho_k^2-\rho_{k+1}^2)^3}+\nonumber\\
&&+\frac{\rho_k^2(\rho_k^8+2\rho_k^6\rho_{k+1}^2-6\rho_k^4\rho_{k-1}^2\rho_{k+1}^2+2\rho_k^2\rho_{k-1}^4\rho_{k+1}^2+\rho_{k-1}^4\rho_{k+1}^4)q_{k-1}^2q_k^4}{(\rho_k^2-\rho_{k-1}^2)^3(\rho_k^2-\rho_{k+1}^2)^4}-\nonumber\\
&&-\frac{\rho_k^2\rho_{k+1}^2(3\rho_k^6+\rho_k^2\rho_{k-1}^2\rho_{k+2}^2+\rho_{k-1}^2\rho_{k+1}^2\rho_{k+2}^2-\rho_k^4(\rho_{k+1}^2+2\rho_{k-1}^2+2\rho_{k+2}^2))q_{k+1}^2q_k^2q_{k-1}^2}{(\rho_k^2-\rho_{k-1}^2)^2(\rho_k^2-\rho_{k+1}^2)^3(\rho_{k}^2-\rho_{k+2}^2)} +\nonumber
\eea\bea
&& + \frac{\rho_k^4\rho_{k+1}^2q_k^2q_{k+1}^4}{(\rho_k^2-\rho_{k+1}^2)^3(\rho_k^2-\rho_{k+2}^2)^2}+ \frac{\rho_k^2\rho_{k+1}^2(-2\rho_k^4+2\rho_{k+1}^2\rho_{k+2}^2+\rho_k^2\rho_{k+2}^2-\rho_k^2\rho_{k+1}^2)q_k^4q_{k+1}^2}{(\rho_k^2-\rho_{k+1}^2)^4(\rho_k^2-\rho_{k+2}^2)^2}+\nonumber \\
&& + \frac{\rho_k^2\rho_{k+1}^2\rho_{k+2}^2q_k^2q_{k+1}^2q_{k+2}^2}{(\rho_k^2-\rho_{k+1}^2)^2(\rho_k^2-\rho_{k+3}^2)^2(\rho_k^2-\rho_{k+3}^2)} + \frac{\rho_k^2\rho_{k+1}^2(\rho_k^2+\rho_{k+1}^2)q_k^6}{(\rho_k^2-\rho_{k+1}^2)^5}
\end{eqnarray}

\section*{Appendix B}

Here we present the results for the functions $f_4$ and $f_6$
calculated in some simple Type I models.
\begin{itemize}
\item[(i)] With only $N=2$ pair of fields $\tilde\phi_i$ and $\phi_i$ 
there is only one coupling and one mass ratio $\rho$. One immediately 
obtains
$f_4=(1+\rho^2)/(2(\rho^2-1)^2)-\rho^2(\ln \rho^2)/(\rho^2-1)^3>0$.
\item[(ii)] Let us consider equally spaced masses squared, 
$\rho_k^2=1+k\delta$. 
In the limit $\delta\to0$ we obtain
$f_4=(1/12)\left[q_1^4+q_N^4+\sum_{k=1}^{N-1}(q_k^2-q_{k+1}^2)^2 \right]>0$.
\item[(iii)] We assume 3 pairs of  fields $\tilde\phi_i$ and $\phi_i$ 
with the most general mass and coupling structure,
$\rho_1=1$, $\rho_2=\rho$, $\rho_3=\tilde\rho$, $q_1=1$ and $q_2=q$, 
generalizing the model considered in \cite{Shih}. 
We obtain
\begin{eqnarray}
f_4 &=& \frac{1+\rho^2}{2(\rho^2-1)^2}+\frac{\rho^2q}{(\rho^2-1)(\rho^2-\tilde\rho^2)} +\frac{(\rho^2+\tilde\rho^2)q^2}{2(\rho^2-\tilde\rho^2)^2} + \nonumber\\
&&+\left[\frac{\rho^2\tilde\rho^2q}{(\tilde\rho^2-1)(\rho^2-\tilde\rho^2)^2}+\frac{\rho^2\tilde\rho^2q^2}{(\rho^2-\tilde\rho^2)^3}\right]\ln \tilde\rho^2- \nonumber \\
&&- \left[\frac{\rho^2}{(\rho^2-1)^3}+\frac{\rho^2(\rho^4-\tilde\rho^2)q}{(\rho^2-1)^2(\rho^2-\tilde\rho^2)^2}+\frac{\rho^2\tilde\rho^2q^2}{(\rho^2-\tilde\rho^2)^3}\right]\ln \rho^2 
\label{ex3}
\end{eqnarray}
The sign of the expression (\ref{ex3}) is shown in Figure \ref{fex3} 
as
a function of $\rho$ and $\tilde\rho$.
\begin{figure}
\begin{center}
\includegraphics[height=5cm]{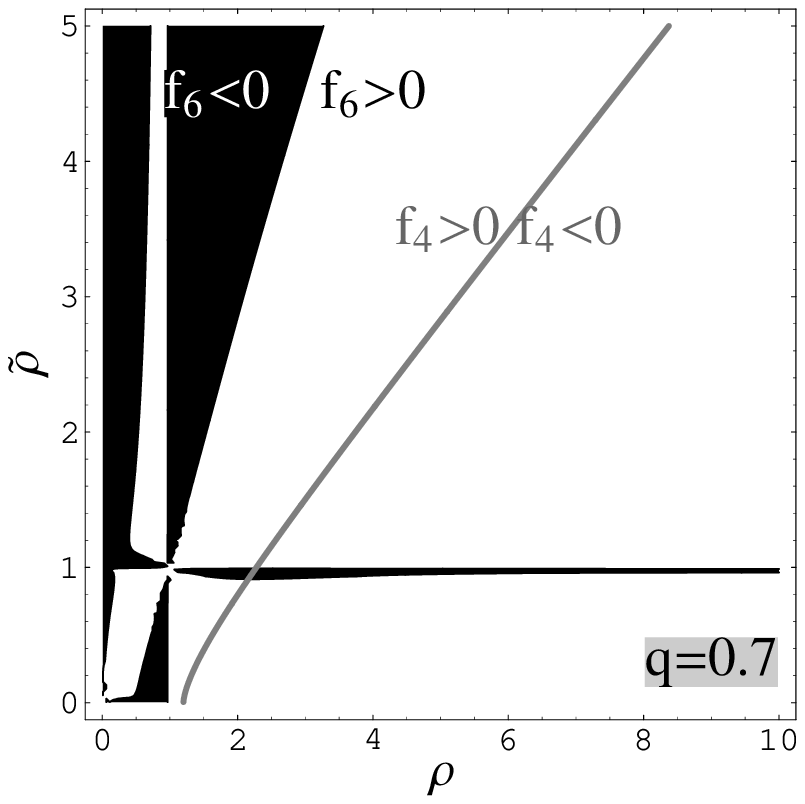}
\hspace{0.5cm}
\includegraphics[height=5cm]{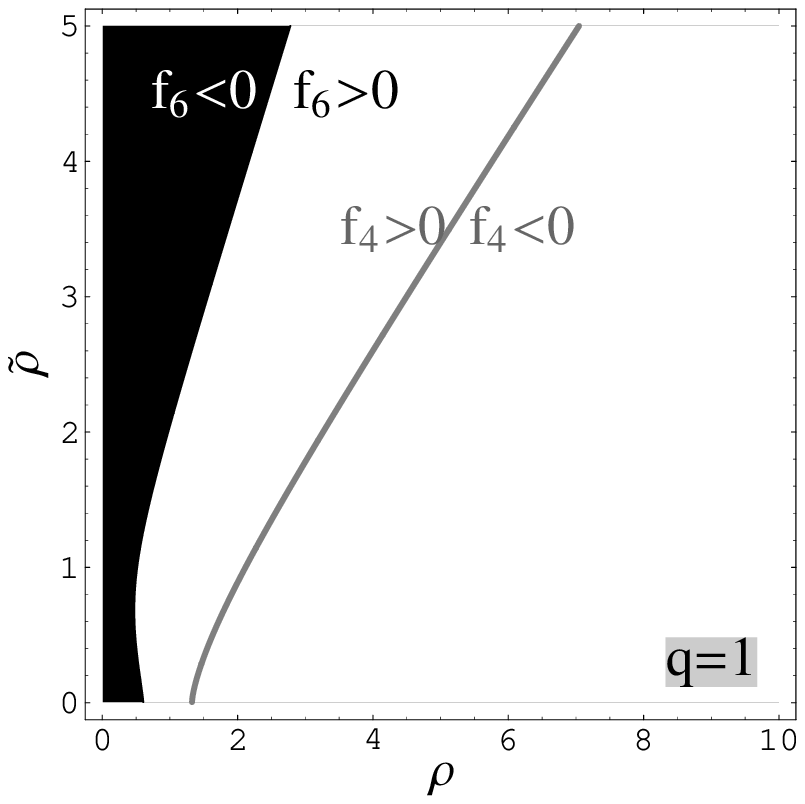}
\hspace{0.5cm}
\includegraphics[height=5cm]{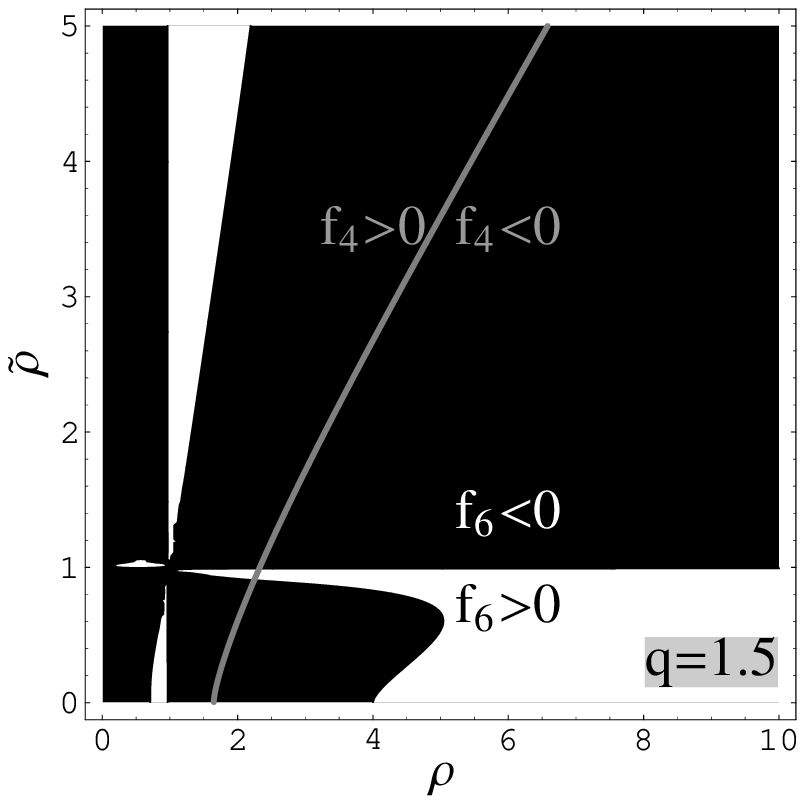}
\end{center}
\caption{\em For the model discusses in Example (iii), 
we plot the sign of the coefficient $f_4$, given in (\ref{ex3}), 
and the coefficient $f_6$ as functions of the mass ratios $\rho$ 
and $\tilde\rho$ for three values of the model parameter $q$. The
black (white) region corresponds to $f_6<0$ ($f_6>0$). 
Regions to the left (right) of the gray line correspond to 
$f_4>0$ ($f_4<0$).  \label{fex3}}
\end{figure}
\item[(iv)] 
We assume $N$ pairs of  fields $\tilde\phi_i$ and $\phi_i$. 
We randomize their relative couplings $q_k$ in the range 
$[1,q_\mathrm{max}]$ with a constant probability density for $\ln q_k$. 
We randomize $\rho_k$ with a constant probability density for $\ln \rho_k$, 
rescaling the results so that  the ratio between the smallest and 
the largest $\rho_k$ is exactly $\rho_\mathrm{max}$. 
We repeat this procedure 100 times to estimate the probability of
the event of interest (either $f_4<0$ and $f_6>0$ or $f_4<0$ and $f_6<0$).
We perform this calculation 20 times to 
assess the variance of this prediction. 
Results showing the dependence $P(f_4<0\,\mathrm{and}\,f_6>0)$ and 
$P(f_4<0\,\mathrm{and}\,f_6<0)$
on $\rho_\mathrm{max}$ for $N=4,\,20$ are shown in Figure \ref{fex4}
as light gray (black) dots for $q_\mathrm{max}=1\,(10)$.
We can conclude, in concordance with Example (iii), that a supersymmetry 
breaking minimum with $X\neq0$ exists in a significant fraction of 
the parameter space, given that the mass hierarchies are sufficiently large. 
This fraction is generically larger for degenerate than nondegenerate 
couplings.
\begin{figure}
\begin{center}
\includegraphics[height=6cm]{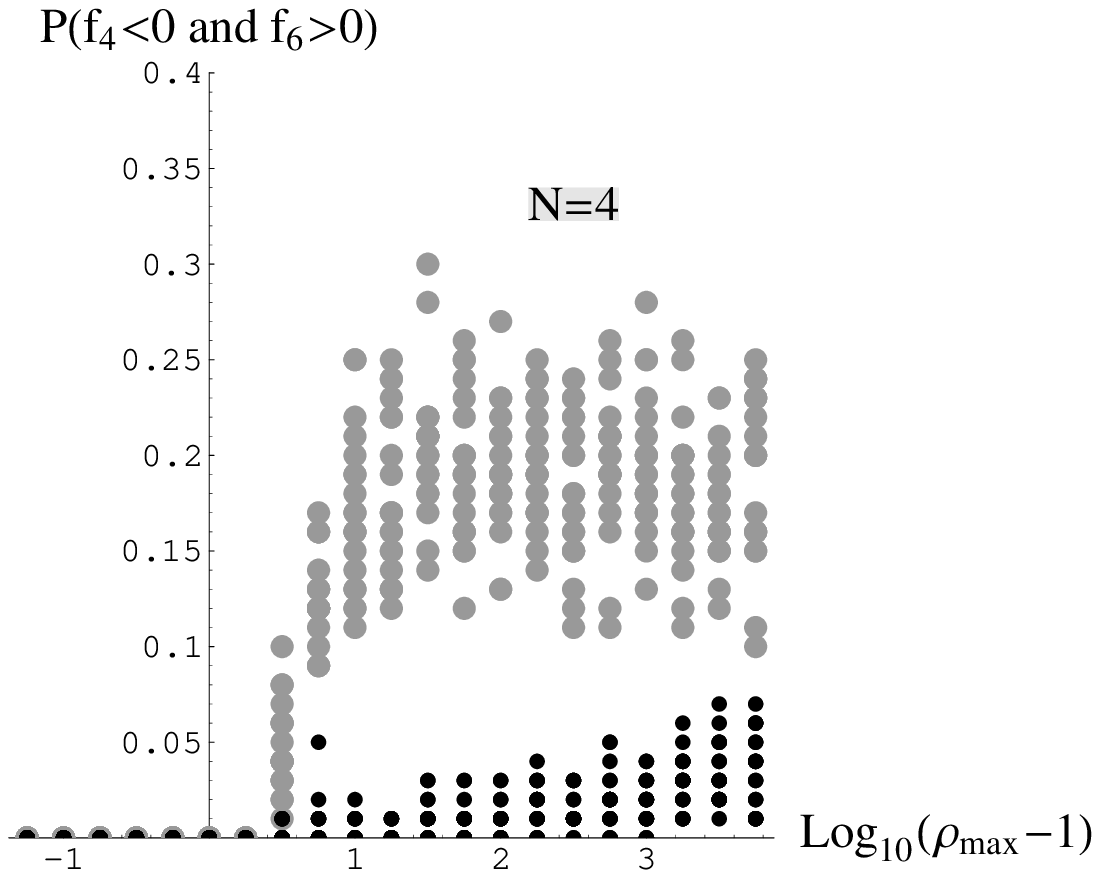}
\hspace{0.5cm}
\includegraphics[height=6cm]{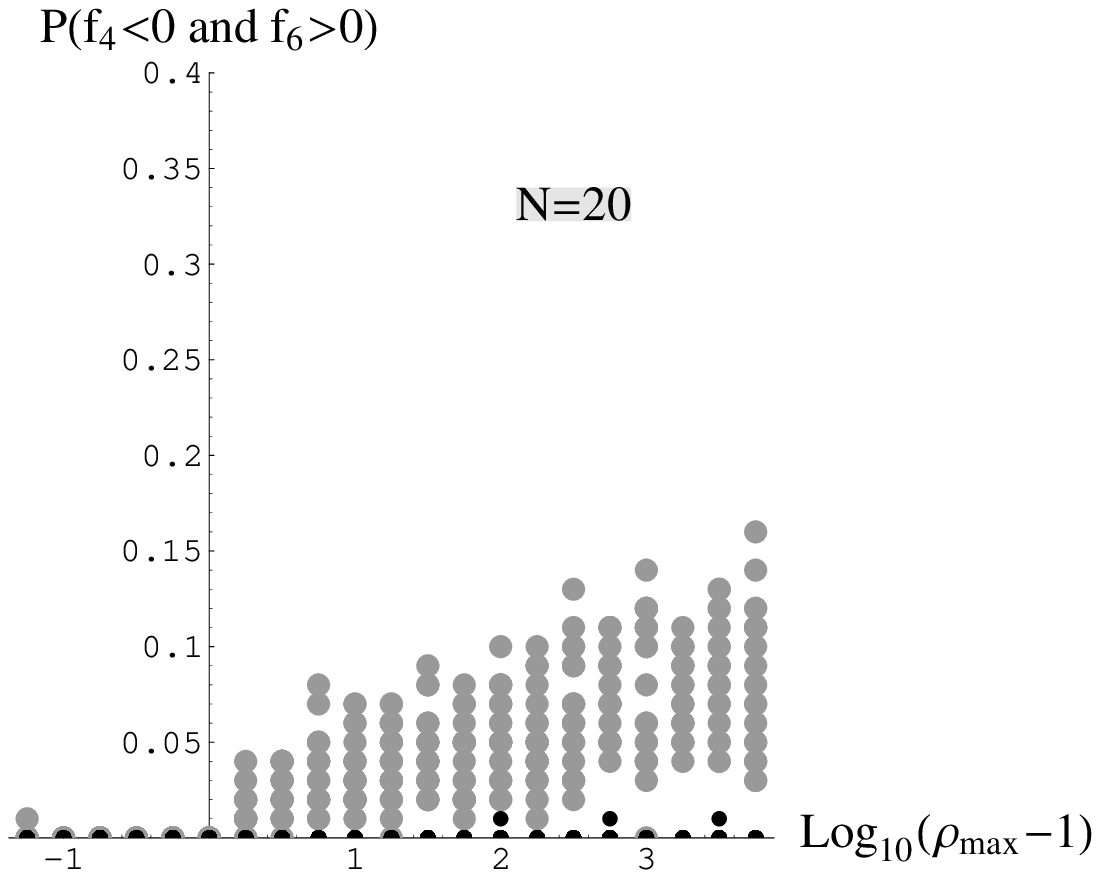}
\\\includegraphics[height=6cm]{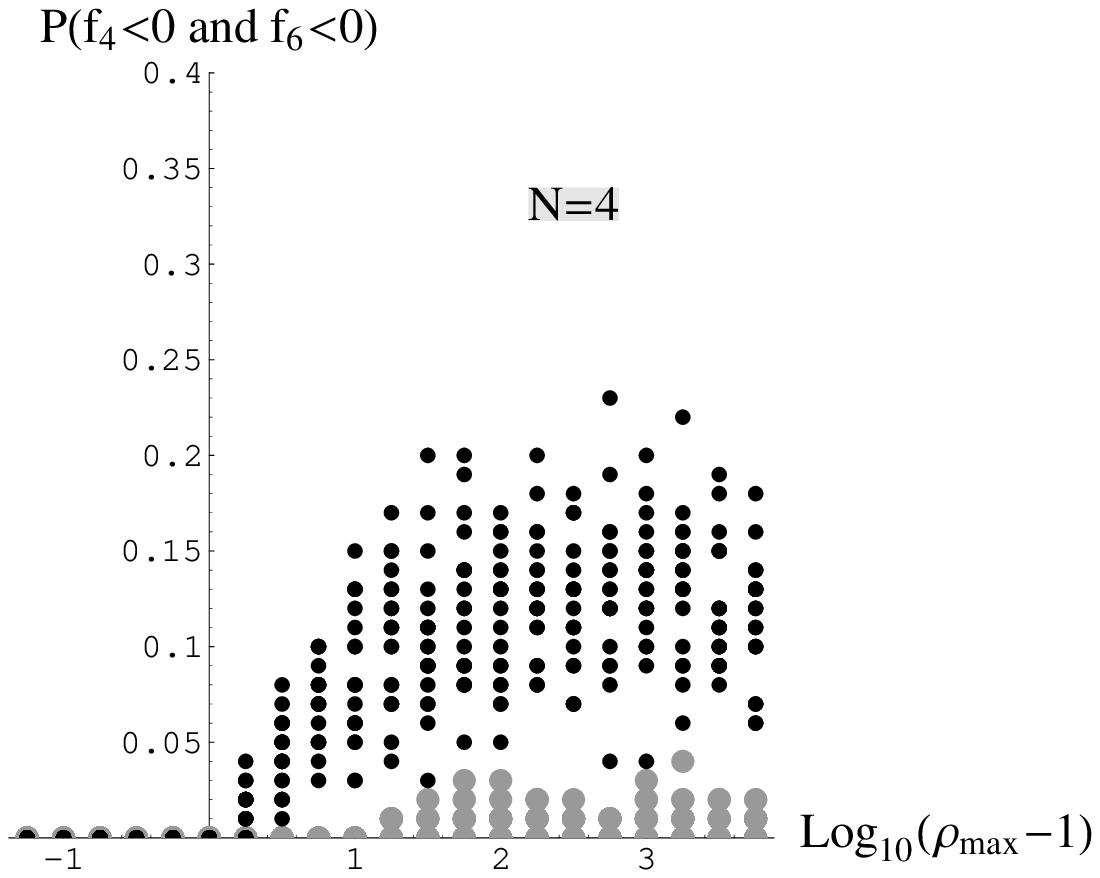}
\hspace{0.5cm}
\includegraphics[height=6cm]{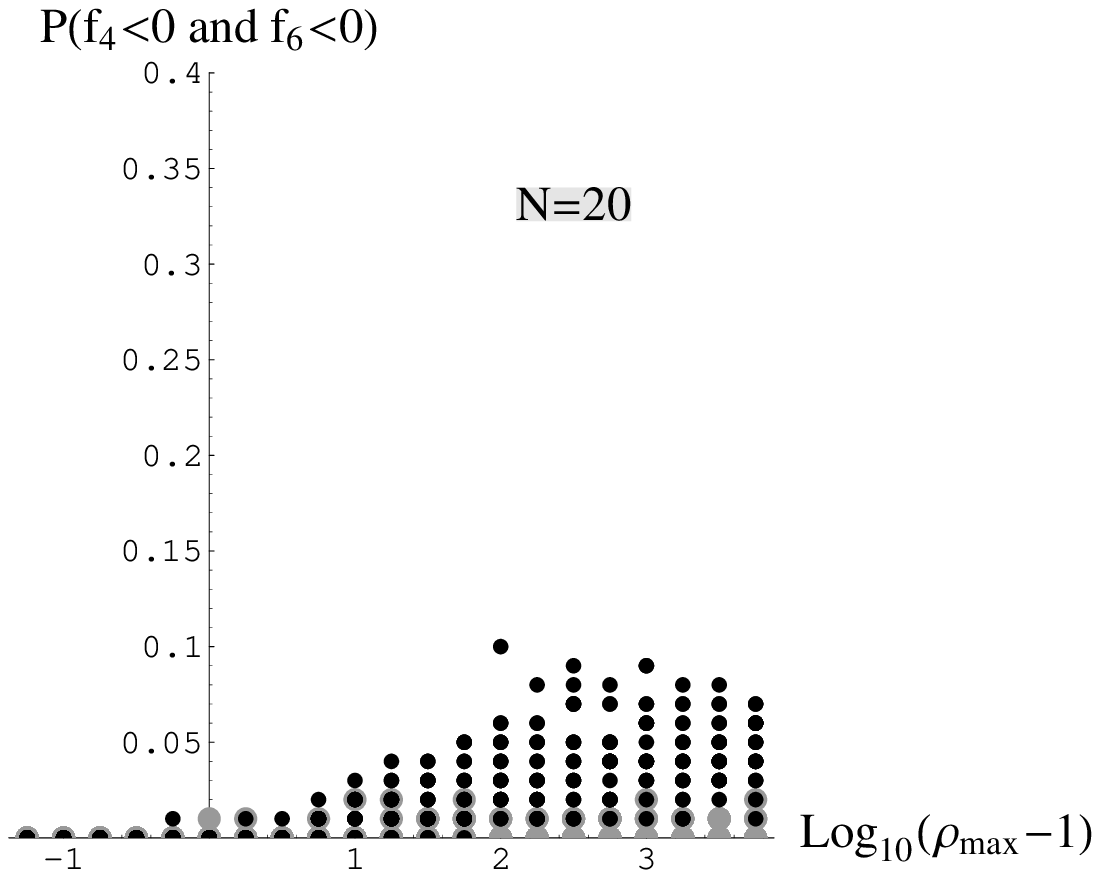}
\end{center}
\caption{\em Results of the numerical procedure employed in Example (iv). 
The plots show $P(f_4<0\,\textrm{and}\,f_6>0)$ and 
$P(f_4<0\,\textrm{and}\,f_6<0)$ for different values of 
$\rho_\mathrm{max}$ for $N=4$ and $N=20$. 
Light gray (black) dots correspond to maximal ratio of couplings 
in the O'Raifeartaigh sector equal to $q_\mathrm{max}=1\,(10)$.  
\label{fex4}}
\end{figure}
\end{itemize}
Below we also discuss some examples of models in which the messenger sector
can affect the position of the supersymmetry breaking minimum.
\begin{itemize}
\item[(v)] In the absence of the O'Raifeartaigh sector and for 
$\tilde N=2$, $ \Mc^s$ is given by 
\begin{equation}
 \Mc^\mathrm{s} = \left( \begin{array}{cc} \lambda_1X & m \\ 0 & \lambda_2X\end{array}\right)
\end{equation}
and we can find the analytic form of eigenvalues of 
$ \Mc^{\mathrm{s}\dagger} \Mc^\mathrm{s}$ 
and numerically solve for a minimum of the effective potential.
In Figure \ref{fex5}a we plot the value of $\sqrt{\lambda_1\lambda_2} X/m$ 
at the minimum as a function of $\sqrt{\lambda_1/\lambda_2}$ 
(and the results agree with those from \cite{EOGM} for this 
parameter equal to 1).
\begin{figure}
\begin{center}
\includegraphics[height=6.5cm]{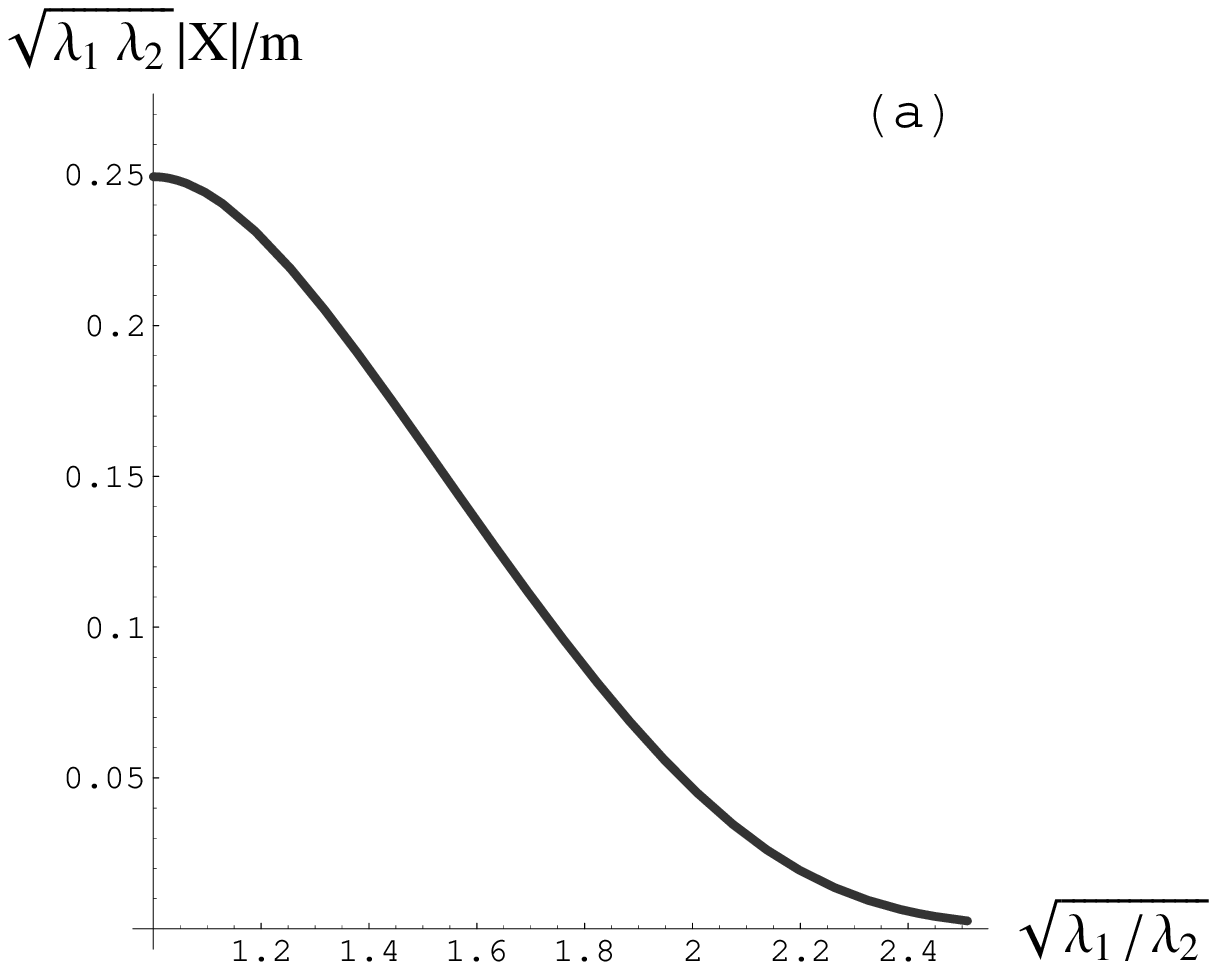}
\hspace{0.5cm}
\includegraphics[height=6cm]{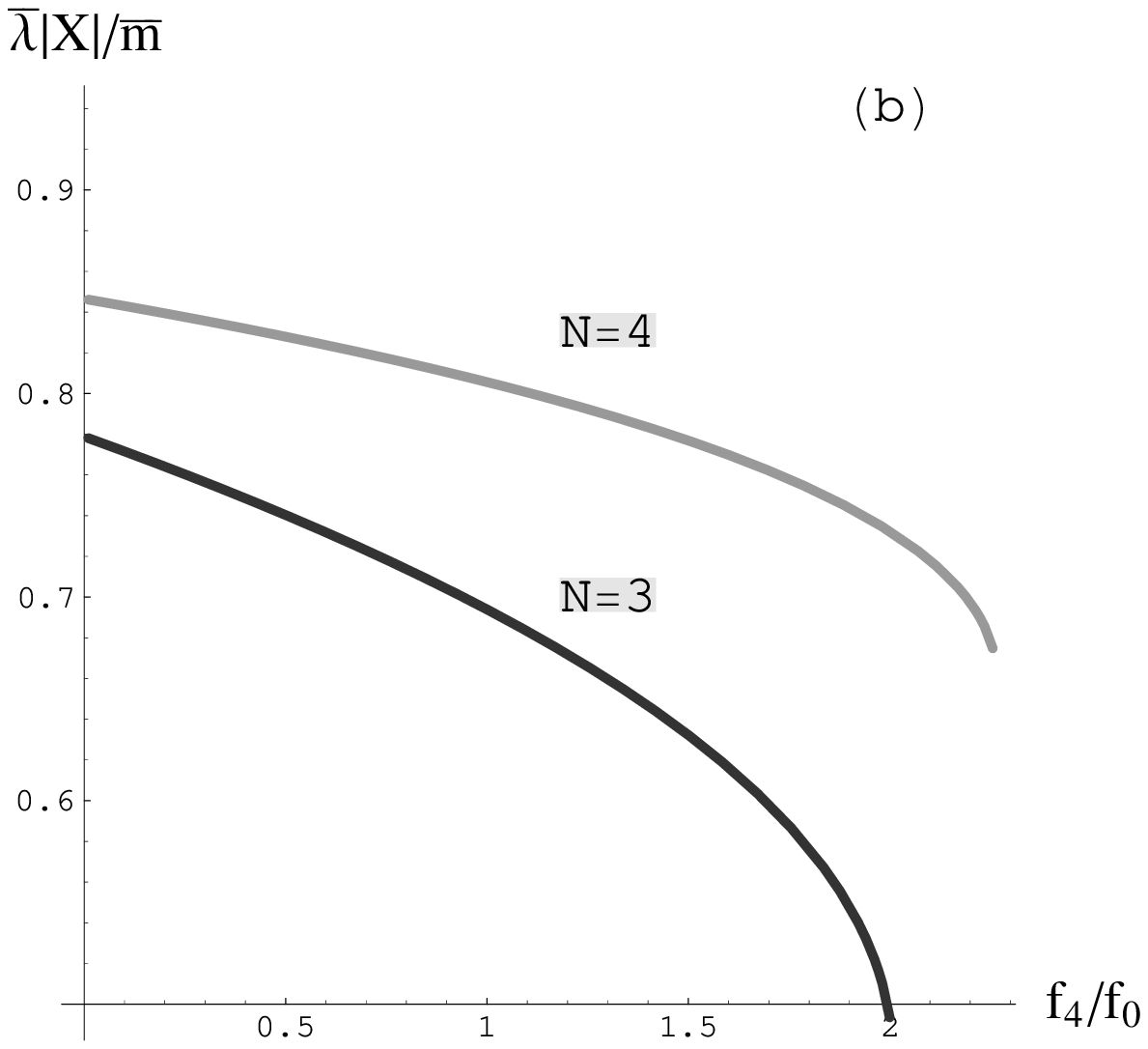}
\end{center}
\caption{\em (a) The location for the supersymmetry breaking minimum
calculated in the absence of supergravity corrections in Example (v);
(b) the location for the supersymmetry breaking minimum
calculated in the absence of supergravity corrections in Example (vi)
for a truncated correction (\ref{eff2d}) to the K\"ahler
potential. \label{fex5}}
\end{figure}
\item[(vi)] For $\tilde N>2$, the eigenvalues of 
$ \Mc^{\mathrm{s}\dagger} \Mc^\mathrm{s}$
have to be computed numerically and $f_4$, $f_0$ are functions of
three or more parameters. We  may therefore employ the expansion 
(\ref{eff2d}) to look for a supersymmetry breaking minimum even for $f_4>0$,
taking $f_4/f_0$ as a free parameter and neglecting the higher order terms. 
The results for of $\bar{\lambda} |X|/\bar{m}$ at the minimum are 
shown in Figure \ref{fex5}b.
Of course, for large values of the expansion parameter
$\bar{\lambda}|X|/\bar{m}$ higher order corrections can become important
and one should treat these results only as qualitative estimates.
\end{itemize}

\section*{Appendix C}

Here we discuss corrections to the expression (\ref{neffdef}) for 
$N_{\mathrm{eff},r}$ originating from messenger mass splittings.

In the Giudice-Rattazzi formalism \cite{GRwave}, the soft masses
of the scalars are extracted from the relation
\beq
m_{\tilde f}^2 (\mu) = - |F|^2 \partial_{\bar X}\partial_X \ln Z_{\tilde f}(\mu,X,X^\dagger)
\label{grscalar}
\eeq
where $Z_{\tilde f}$ is the sfermion wave function renormalization constant
satisfying:
\beq
\frac{\mathrm{d}\ln Z_{\tilde f}}{\mathrm{d}\tau} = \sum_{r=1}^3 C_{\tilde f}^r \frac{g^2_r}{4\pi^2} \, ,
\label{zq}
\eeq
where $\tau=(1/(16\pi^2))\ln(\mu/\Lambda_\mathrm{UV})$.
Let us restrict the discussion to a single gauge group (dropping the index $r$
from now on), as the contributions
from different gauge groups to $\ln Z_{\tilde f}$ add up. 
Equation (\ref{zq}) can be rewritten
using the RGE for the gauge coupling:
\beq
\frac{\mathrm{d}\ln Z_{\tilde f}}{\mathrm{d}\tau} = -\frac{2C_{\tilde f}}{\beta_g} \frac{\mathrm{d}\ln g^{-2}}{\mathrm{d}\tau} \, ,
\eeq
where $\beta_g$ is the MSSM beta function of the gauge coupling.
Taking into account multiple messenger thresholds, we arrive at the result:
\beq
\ln Z_{\tilde f} (\mu,X,X^\dagger)-\ln Z_{\tilde f}(\Lambda_\mathrm{UV}) = -2C_{\tilde f} \sum_{j=0}^{\tilde N} \frac{1}{\beta_g+jS} \left\{ \ln\left[g^{-2}(\Mc_j)\right]-\ln\left[g^{-2}(\Mc_{j+1})\right]\right\} \, ,
\label{zq1}
\eeq
where $S$ is the Dynkin index of the messenger representation ($S=1$ for
a $\mathbf{N}+\mathbf{\bar N}$ pair of $SU(N)$) and $\mathcal{M}_0$
and $\mathcal{M}_{\tilde N+1}$ stand for $\mu$ and $\Lambda_\mathrm{UV}$,
respectively.
Eq.\ (\ref{zq1}) is rather complicated for practical applications. One can
try to simplify the analysis by combining contributions to sfermion masses
originating at each messenger threshold $\mu=\Mc_j$. Choosing
$\mu$ and $\Lambda_\mathrm{UV}$ infinitesimally close
to the threshold, one can rewrite the first line of (\ref{zq1}) as:
\beq
\ln Z_{\tilde f} (\mu,X,X^\dagger)-\ln Z_{\tilde f}(\Lambda_\mathrm{UV}) = -2C_{\tilde f} \left[ \frac{1}{\beta_g} \ln\left(\frac{g^{-2}(\mu)}{g^{-2}(\Mc_j)}\right)+\frac{1}{\beta_g+S} \ln\left(\frac{g^{-2}(\Mc_j)}{g^{-2}(\Lambda_\mathrm{UV})}\right)\right]
\eeq
Expanding this expression in $\ln((\Mc_j)^2/\Lambda^2_\mathrm{UV})$, we find:
\beq
\ln Z_{\tilde f} (\mu,X,X^\dagger)-\ln Z_{\tilde f}(\Lambda_\mathrm{UV}) = C_{\tilde f} S \left( \frac{g^2}{16\pi^2} \right)^2 \left( \ln \frac{\Mc_j^2}{\Lambda^2_\mathrm{UV}} \right)^2 + O\left( \left( \ln \frac{\Mc_j^2}{\Lambda^2_\mathrm{UV}} \right)^3 \right)
\label{zq2}
\eeq
Summing all such results for $j=1,\ldots,\tilde N$ 
and substituting to (\ref{grscalar}), 
we obtain the result (\ref{mass:scalar}).

\begin{figure}
\begin{center}
\includegraphics[height=7cm]{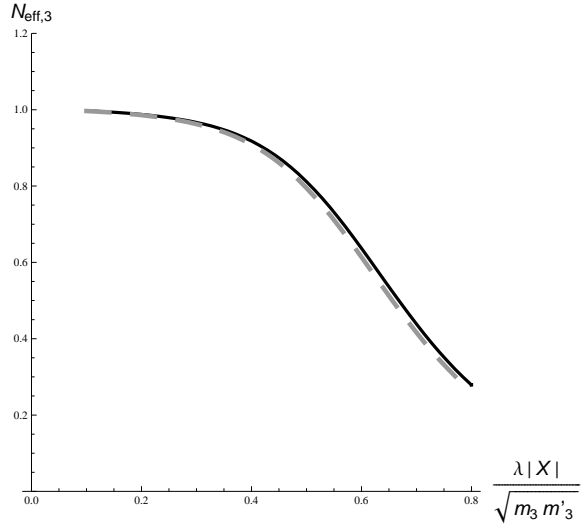}
\end{center}
\caption{\em Values of $N_\mathrm{eff}$ calculated in the $SU(5)$-symmetric
version of the model given in eq.~(\ref{wex10}) in Section \ref{exx}. The black solid line correspond to employing the full Giudice-Rattazzi method as described in Appendix C; the gray dashed line correspond to the simplified result following from (\ref{mass:scalar}). \label{fneff}}
\end{figure}

In this derivation, we eventually ignored the higher powers of
the logarithm, which is legitimate, as, after differentiation with respect
to $X$ and $X^\dagger$, we can set $\Lambda_\mathrm{UV}$ to the scale
of the messenger threshold.
One can, however, worry that in models with large splittings
between the messenger masses,
ignoring the running of the sfermion
masses and gauge couplings between the messenger thresholds
(which the full formula (\ref{zq1}) takes into account up to top quark Yukawa
corrections) can reduce the
accuracy of the result (\ref{mass:scalar}). In order to address this issue
quantitatively, we studied the model presented in eq.~(\ref{wex10a}) in
Section \ref{exx} with $m_3/m_2=9$. 
For simplicity we took the triplet sector in which the mass
splittings are the largest
and, without stabilizing $X$ at the minimum of the potential, we calculated
$N_\mathrm{eff,3}$ using the full Giudice-Rattazzi result (\ref{zq1}) and
the simplified method following from (\ref{mass:scalar}). 
Since the gauge coupling
had to be specified in the former approach, we took $g^2=2/3$ and $\beta_g=-3$.
A comparison of the two calculations is presented in Figure \ref{fneff}. We see
that even with the messenger mass hierarchy larger than $10^3$ in this
model, the simplified result does not deviate
from the more accurate one. A similar calculation with an even larger mass ratio $m_3/m_2=100$
revealed only small discrepancies of the order of $10\%$ between the two approaches.
We therefore conclude that for practical purposes we can safely neglect the threshold corrections
to (\ref{mass:scalar}) coming from split messenger masses. 

\section*{Appendix D}

Here we employ the analytical formulae for the solutions of the
1-loop RGE for the MSSM derived in \cite{bottom-up} to discuss
the radiative breaking of the electroweak symmetry and the predictions for
$\mu$ and $B$ for different choices of the effective messenger numbers
$N_{\mathrm{eff},r}$.

The following equations describe the minimization conditions of the
higgs sector of the scalar potential of the MSSM:
\bea
\label{ewsb1}
\mu^2 &\simeq& -m_{H_2}^2(\tau) \\
 B(\tau)\mu\tan\beta &\simeq& m_{H_1}^2(\tau)-m_{H_2}^2(\tau)
\label{ewsb2}
\eea
The running mass parameters are presented as functions of
$\tau=\frac{1}{(4\pi)^2}\ln\frac{M_\mathrm{low}}{\mathcal{M}_\mathrm{mess}}$.
Here we neglected the running of $\mu$, terms proportional to $M_Z^2$ 
and $1/\tan^2\beta$. We also assumed that the scale $M_\mathrm{low}$
at which (\ref{ewsb1}) and (\ref{ewsb2}) are evaluated
is chosen so that the threshold $y_t^4$-proportional 
corrections to the masses and couplings
relevant for EWSB vanish, i.e.~$M_\mathrm{low}$ is the geometric mean of
the stop masses.

The parameters $m_{H_2}^2(\tau)$, $m_{H_1}^2(\tau)$ and $B(\tau)$
can be expressed in terms of the input values
$m_{H_2}^2(0)$, $m_{H_1}^2(0)$ and $B(0)$
at the scale $\mathcal{M}_\mathrm{mess}$: 
\bea
m_{H_2}^2(\tau) &=& \left(1-\frac{y}{2}\right)m_{H_2}^2(0)-\frac{y}{2}\left( m_Q^2(0)+m_U^2(0)\right)+\left(\eta_{H_2}+\frac{y}{2}(-\hat\eta+y\hat\xi^2)\right) M_{1/2}^2 + \nonumber \\ &&+\frac{1}{22}\left(\frac{g_1^2(\tau)}{g_1^2(0)}-1\right)\left(m_{H_2}^2(0)-m_{H_1}^2(0)\right) \label{mhu} \\
m_{H_1}^2(\tau) &=& m_{H_1}^2(0)+\eta_{H_2}M_{1/2}^2-\frac{1}{22}\left(\frac{g_1^2(\tau)}{g_1^2(0)}-1\right)\left(m_{H_2}^2(0)-m_{H_1}^2(0)\right) \\
B(\tau) &=& B(0)+\left(-\xi_B+\frac{y}{2}\hat\xi\right)M_{1/2} \, .
\label{bsol}
\eea
Here, $y=y_t^2/(y_t^\mathrm{FP})^2$, where $y_t$ is the running top Yukawa
coupling and $y_t^\mathrm{FP}$ is its quasi-fixed point value.
$M_{1/2}$ is the gluino mass at $\Mc_\mathrm{mess}$.
The coefficients $\xi_B$, $\eta_{H_2}$, $\hat\xi$ and $\hat\eta$ 
in (\ref{mhu})-(\ref{bsol})
depend on the gauge couplings, gaugino mass ratios and some
group-theoretical factors; they are defined in \cite{bottom-up}.
In Table \ref{tabd}, we present their values for two choices of 
$\Mc_\mathrm{mess}=2\cdot 10^{15}\,\mathrm{GeV}$
and $\Mc_\mathrm{mess}=2\cdot 10^{5}\,\mathrm{GeV}$, discussed in
Section 5.
Since their values of these coefficients depend on threshold corrections
to the gauge couplings (mainly to the strong coupling, 
$\hat\alpha_s=\alpha_s/(1-\Delta\alpha_s)$) and the value of 
$M_\mathrm{low}$ at which we stop the RG
evolution to study the electroweak symmetry breaking and to calculate the
mass spectra of the supersymmetric particles, we show the results
for $\alpha_s=0.118$ and for different values of $\Delta\alpha_s$ and 
$M_\mathrm{low}$.

\begin{table}
\begin{center}
\begin{tabular}{|c|rccc|ccc|}
\hline
 & \multicolumn{4}{|c|}{$\Mc_\mathrm{mess}=2\times 10^{15}\,\mathrm{GeV}$} & \multicolumn{3}{|c|}{$\Mc_\mathrm{mess}=2\times 10^5\,\mathrm{GeV}$} \\
\hline
 & $M_\mathrm{low}= $& $M_Z$ & $500\,\mathrm{GeV}$ &  $1000\,\mathrm{GeV}$ & $M_Z$ & $500\,\mathrm{GeV}$ &  $1000\,\mathrm{GeV}$ \\
\hline
$\Delta\alpha_3=0$ & $(y_t^\mathrm{FP})^2=$ & 1.30 & 1.26 & 1.25 & 2.52 & 2.94 & 3.21 \\
& $\xi_B=$ & 0.508 & 0.485 & 0.476 & 0.056 & 0.044 & 0.039 \\
& $\eta_{H_2}=$ & 0.427 & 0.409 & 0.402 & 0.023 & 0.018 & 0.016 \\
& $\hat\xi=$ & 2.01 & 1.76 & 1.67 & 0.423 & 0.302 & 0.258 \\
& $\hat\eta=$ & 11.3 & 9.00 & 8.25 & 1.21 & 0.778 & 0.693 \\
\hline
$\Delta\alpha_3=-0.10$ & $(y_t^\mathrm{FP})^2=$ & 1.24 & 1.21 & 1.20 & 2.46 & 2.89 & 3.16 \\
& $\xi_B=$ & 0.527 & 0.503 & 0.493 & 0.060 & 0.047 & 0.042 \\
& $\eta_{H_2}=$ & 0.459 & 0.440 & 0.432 & 0.026 & 0.021 & 0.018 \\
& $\hat\xi=$ & 1.88 & 1.68 & 1.58 & 0.390 & 0.280 & 0.241 \\
& $\hat\eta=$ & 9.92 & 8.07 & 7.43 & 1.08 & 0.705 & 0.583 \\
\hline
$\Delta\alpha_3=-0.20$ & $(y_t^\mathrm{FP})^2=$ & 1.19 & 1.17 & 1.16 & 2.41 & 2.85 & 3.12 \\
& $\xi_B=$ & 0.545 & 0.521 & 0.511 & 0.064 & 0.050 & 0.045 \\
& $\eta_{H_2}=$ & 0.491 & 0.471 & 0.463 & 0.030 & 0.024 & 0.021 \\
& $\hat\xi=$ & 1.77 & 1.57 & 1.50 & 0.363 & 0.263 & 0.226 \\
& $\hat\eta=$ & 8.89 & 7.33 & 6.79 & 0.974 & 0.647 & 0.537 \\
\hline
\end{tabular}
\end{center}
\caption{\em Coefficients entering the solutions (\ref{mhu})-(\ref{bsol}) of the RG equations. \label{tabd}}
\end{table}

Now we would like to use the solutions (\ref{mhu})-(\ref{bsol}) 
to determine which classes of the initial conditions at the scale
$\Mc_\mathrm{mess}$ correspond to successful breaking of the
electroweak symmetry with small $\mu$ and $B(0)$.
To this end, we define $\Xi'=\mu^2/M_{1/2}^2$
and rewrite (\ref{ewsb1}) as
\bea
\Xi_{H_2}-\Xi' &=& \frac{1}{N_\mathrm{eff,2}}\left[ \frac{3}{2}(1-y)\frac{g_2^4(0)}{g_3^4(0)}+\left(\frac{9}{50}-\frac{13}{50}y\right)\frac{g_1^4(0)}{g_3^4(0)} \right] + \nonumber\\
&&+\frac{1}{N_\mathrm{eff,3}}\left[ -\frac{8}{3}y + \left(\frac{3}{25}-\frac{13}{75}y\right)\frac{g_1^4(0)}{g_3^4(0)}\right] \, ,
\label{mucon}
\eea
where $\Xi_{H_2}=-\eta_{H_2}+\frac{y}{2}(\hat\eta-y\hat\xi^2)$.
This equation simplifies considerably if $\Mc_\mathrm{mess}$ 
is close to the unification scale,
since we can assume that the gauge couplings are approximately equal:
\beq
\Xi_{H_2}-\Xi' = \frac{1}{N_\mathrm{eff,2}}\left( \frac{42}{25}-\frac{101}{50}y\right) + \frac{1}{N_\mathrm{eff,3}}\left( -\frac{8}{3}+\frac{19}{50}y\right) \, .
\eeq
The left-hand side of this equation is positive and of the order of a few.
The coefficient of $1/N_\mathrm{eff,2}$ is small and 
positive, while
the coefficient of $1/N_\mathrm{eff,3}$ is negative and of the order of unity.
Thus, we can conclude that solutions with small $\mu\ll M_{1/2}$, 
which correspond
to $0<\Xi'\ll 1$, will require a small $N_\mathrm{eff,2}\ll N_\mathrm{eff,3}$.
Eq.~(\ref{mucon}) also simplifies for $\Mc_\mathrm{mess}$ much smaller than the
unification scale, when we can neglect the terms proportional to $g_1^4(0)$
and take $1-y\approx 1$.
We then obtain
\beq
\Xi_{H_2}-\Xi' = \frac{1}{N_\mathrm{eff,2}} \frac{3}{2}\frac{g_2^4(0)}{g_3^4(0)} - \frac{1}{N_\mathrm{eff,3}} \frac{8}{3}y \, .
\eeq
Taking $\Mc_\mathrm{mess}=2\times10^5$
as a particular example, we see that $\Xi_{H_2}$ approximately vanishes,
so the two terms on the right-hand side must cancel out:
\beq
N_\mathrm{eff,3} \simeq N_\mathrm{eff,2} \frac{16y}{9}\frac{g_2^4(0)}{g_3^4(0)} 
\eeq
The numerical coefficient is $\approx 3\approx 10^{0.4}$, so this equation is a
reasonable approximation of the results in Figure \ref{fewsb}.

We see that in both examples discussed above, one can obtain solutions
with small $\mu$, given that there is an appropriate hierarchy 
$N_\mathrm{eff,2}\ll N_\mathrm{eff,3}$. 
Due to this requirement, there are not any solutions
corresponding to $N_\mathrm{eff,2}=N_\mathrm{eff,3}$ which is the
outcome of conventional models with gauge mediated supersymmetry breaking.

Having found the conditions for the existence of the solutions with
small $\mu$, we would like to know if some of these solutions correspond
to small $B^{(0)}=\pm\mu/\sqrt{3}$, as in (\ref{gmm1}) and (\ref{gmm2}). 
In order to determine it, we can rewrite (\ref{ewsb2}) as:
\beq
\left(\pm\frac{1}{\sqrt{3}}\Xi'+\Xi_B\right)\Xi'\tan\beta \simeq \nu \, ,
\label{eqb}
\eeq
where
\beq
\nu=\frac{1}{N_\mathrm{eff,2}}\left(\frac{3}{2}\frac{g_2^4(0)}{g_3^4(0)}+\frac{9}{50}\frac{g_1^4(0)}{g_3^4(0)}\right)
+\frac{1}{N_\mathrm{eff,3}} \frac{3}{25}\frac{g_1^4(0)}{g_3^4(0)} + \eta_{H_2}
\eeq
and $\Xi_B=-\xi_B+y\hat\xi/2$.
The solutions with small $\mu$ correspond
to $N_\mathrm{eff,2}\ll N_\mathrm{eff,3}$, so we can write
$\nu\approx \gamma/N_{\mathrm{eff,2}}+\eta_{H_2}\simgt O(1)$
with the coefficient $\gamma$ changing from $3/2$ to $27/25$ for
$\Mc_\mathrm{mess}$ inreasing from $2\cdot10^5\,\mathrm{GeV}$
to $2\cdot10^{15}\,\mathrm{GeV}$.
Equation (\ref{eqb}) has a solution with $0<\Xi'\ll 1$, if
$\nu/\tan\beta$ is a sufficiently small number; this can be obtained
by choosing an appropriate value of $\tan\beta$, unless $N_{\mathrm{eff},2}$ is very small.

To summarize, we can obtain solutions
with small $B=|\mu|/\sqrt{3}\ll M_{1/2}$ 
for both low and high scale of the gauge mediated
symmetry breaking, given that appropriate hierarchy between 
$N_\mathrm{eff,2}$ and $N_\mathrm{eff,3}\gg N_\mathrm{eff,2}$
is arranged for and that
$N_\mathrm{eff,2}$ is not too small. 
These solutions require doublet-triplet splitting in the messenger sector;
a much stronger splitting is necessary for solutions with a high scale
$\Mc_\mathrm{mess}$ for which the Giudice-Masiero mechanism can be made work.

\end{document}